\documentclass[10pt,final,twocolumn,twoside]{IEEEtran}
 \usepackage{setspace}
 \usepackage{calligra}
 \usepackage{latexsym}
 \usepackage{times}
 \usepackage{graphicx}
 \usepackage{color}
 \usepackage{algorithmic}
 \usepackage{algorithm}
 \usepackage{amssymb,amsmath}
 \usepackage{multirow}
 \usepackage{colortbl} 
\usepackage{cite}
 \usepackage[table]{xcolor}
\usepackage[bookmarksnumbered=true,bookmarks,backref]{hyperref}
 \usepackage{float}
 \usepackage{subfigure}

\DeclareMathOperator*{\argmin}{arg\,min}
\DeclareMathOperator*{\argmax}{arg\,max}
\newcommand{\mbf}[1]{\mathbf{#1}}
\newcommand{\mbs}[1]{\boldsymbol{#1}}
 
 \graphicspath{{../figures/}}
\begin{document}
\author{Satyam Dwivedi, Alessio De Angelis, Dave Zachariah, Peter H\"{a}ndel}
%
\title{Joint Ranging and  Clock Parameter Estimation by Wireless Round Trip
  Time Measurements}
\maketitle

\begin{abstract}
In this paper we develop a new technique for estimating fine clock errors and
range between two nodes simultaneously by two-way time-of-arrival
measurements using impulse-radio ultra-wideband signals. Estimators
for clock parameters and the range are proposed that are robust with
respect to outliers. They are analyzed numerically and by means of
experimental measurement campaigns. The technique and derived
estimators achieve accuracies below $1$\,Hz for frequency estimation,
below   $1$\,ns for phase estimation and $20$\,cm for range
estimation, at $4$\,m distance using $100$\,MHz clocks at both nodes. 
\textcolor{black}{
Therefore, we show that the proposed joint approach is practical and can simultaneously provide clock synchronization and positioning in an experimental system.}  
\end{abstract}
\vspace{-0.15in}

\section{Introduction}

A common time reference among nodes along with node positioning
enables coordination in time and space domains. By such coordination
the nodes can execute tasks efficiently in internet-of-things (IOT),
machine-to-machine (M2M) communication, or similar scenarios
\cite{gubbi2013internet,M2M1,M2M2}. In certain scenarios time synchronization among nodes is
necessary for positioning. Specifically, it is needed in positioning techniques where
time of arrival (TOA) measurements are required and measurements are
performed using a clock. As an example, in the commercially available Ubisense system, sensor
nodes are synchronized using cable in order to perform time difference of
arrival (TDOA) measurements of ultra-wideband (UWB) pulses from tag nodes
\cite{ubisense_ref}.  In these types of systems, replacing the cabling with
wireless synchronization will reduce the installation cost in stationary
environments and increase in flexibility will open up new mobile
applications. Positioning and effects of clock parameter mismatch have been addressed jointly before in
\cite{zheng_wu,DBP_joint,TDOA_clock_Gholami} and the references 
therein.  In \cite{zheng_wu}, optimal algorithms were derived for
joint positioning and synchronization with anchor uncertainty in position and
time. In \cite{DBP_joint}, the authors proposed distributed algorithms for
positioning and synchronization in UWB \textit{ad-hoc} networks. 
In \cite{TDOA_clock_Gholami}, the effect of clock skew was discussed and
estimators for TDOA-based positioning were formulated. 
A few
recent works on joint ranging and clock synchronization are mentioned in \cite{chepuri_2012, chepuri_2013, etzlinger_2014,
carroll_2014}. In \cite{chepuri_2013}, authors have proposed a network
wide clock synchronization and ranging mechanism by pairwise
timestamp measurements aided with passive listening. In
\cite{etzlinger_2014}, authors have proposed clock synchronization and localization
 along with a low cost testbed for its demonstration. 

Our main contribution in this paper lies in suggesting a
measurement mechanism which provides information about
clock parameters and range between nodes. Further, we propose
joint estimators to precisely estimate range and clock parameters from the
measurements. The estimated range can subsequently be used for positioning
using standard techniques \cite{henk_positioning1, molisch_pos_UWB,  schedule_comm_lett}. 
Such an approach would simplify physical
infrastructure and obviate the need for e.g. laying cables for
synchronization in positioning systems. 

Similarly, wireless clock synchronization can be useful in several other
applications; which either presently rely on wired synchronization or
work asynchronously but exhibit performance loss in the absence of
synchronization.  For instance, many packet-based radio systems
operate asynchronously and suffer packet collisions. Joint positioning
and time synchronization can help reduce such losses in a network. 
Traditionally, time synchronization is a necessity for a variety of networks. 
Network time protocol (NTP) is a well-known mechanism to synchronize clocks over the
internet \cite{NTP_paper}. Similarly, time synchronization protocol
for sensor networks (TPSN) is in use \cite{ganeriwal2003timing}.   In
the reference broadcast synchronization (RBS) protocol, time synchronization is
achieved by broadcast beacon and receiver timestamp exchanges
\cite{elson2002fine}. In all these network synchronization mechanisms,
the clocks are eventually synchronized between two nodes in a
network. In this paper, by contrast, we avoid timestamp exchanges,
and hence communication overhead, using round-trip time
 (RTT) measurements. This enables simultaneous estimation of the clock
 parameters and range between two nodes.

To achieve high accuracy in wireless clock synchronization it is
necessary to estimate the delay arising from the signal propagation
time between nodes. Here we consider using UWB as the wireless
technology and demonstrate results using a UWB-based radio
with a very accurate time measuring device or chronometer. Our
previous work has demonstrated time measurement precision up to
$50$\,ps using a time-to-digital
converter (TDC) \cite{DeAngelisEtAl2013}. Furthermore, usage of UWB in very accurate range and
position estimation is already well known 
\cite{henk_positioning1, molisch_pos_UWB, DBP_joint, Cui_scholtz}.  
Under line of sight conditions the signal propagation delay is directly proportional to
the distance between the nodes. Estimating this delay therefore
enables ranging as a by product of clock synchronization.

Very few clock parameter estimation results have been verified
experimentally in the literature. In
\cite{elson2002fine}, measurements were captured using Berkeley
motes and logic analyzers and verified offline by synchronization
algorithms. The timescale and accuracy in these experiments
 are in microseconds. Whereas, in our proposed system the timescale
 and accuracy is in the order of a  nano-second, as shown in
 subsequent sections. In \cite{carbone_perugia}, a timestamping system for temporal
 information dissemination is proposed achieving sub-nanosecond
 accuracy with UWB pulses. The timestamping system is
 integrated in a ZigBee platform for wireless sensor networks. However, in
 \cite{carbone_perugia}, one-way time-of-arrival (TOA) is considered, and
 range estimation is not supported, differently from our proposed
 technique.

In the literature various theoretical models and methods have been
discussed for clock synchronization over wireless networks, cf. \cite{Serpedin_YUW, Bernhard_Henk,Ferris_Kumar}. 
In \cite{Bernhard_Henk},
authors have suggested message-passing methods for network clock
estimation. The relative clock model
between two clocks adopted in next section is discussed in
\cite{Ferris_Kumar} in greater detail. In \cite{Isaac_handel}, the
Cram\'{e}r-Rao bounds were derived on the parameters for this clock model. 
In subsequent sections we will
propose the usage of a clocked delay in the slave node. In our previous work, the generated delay is assumed to be fixed
and analog \cite{schedule_comm_lett,
  schedule_self,schedule_eurasip}. Here, by contrast, we turn the clock
into a delay-generating device producing informative measurements. 

Since the proposed technique estimates the range between a
master and a slave node, it can be used as a fundamental building
block of scalable and cooperative distance-based techniques for
network positioning. Specifically, an example of a widely-used
technique exploiting node-to-node ranging is based on
multi-dimensional scaling (MDS)
\cite{shang2003localization}. Furthermore, cooperative message-passing
methods based on Bayesian networks have been developed in
\cite{henk_positioning1}, and experimentally evaluated on a UWB
ranging platform in \cite{experiment_win_dardari}. An analysis and implementation of such methods is beyond of the scope of the present paper.

The remainder of the paper is organized as follows: Section \ref{sec:sys_model}
describes the considered system model for the paper. In section
\ref{sec:mechanism}, the clock parameter measurement mechanism is
introduced. This section also introduces the measurement model. 
In section \ref{sec:estim}, estimators are proposed for
estimating clock parameters and range from the measurements.
Section \ref{sec:numerical} is composed of a numerical evaluation of the model
proposed in section \ref{sec:sys_model}. Section \ref{sec:experiments} provides
results from experimental setup and performance of estimators on the
acquired data. In section \ref{sec:conclusion} we conclude the paper.

\emph{Notation:} $\mbf{1}$ denotes a column vector of $1$s. $\mbf{a} \odot
\mbf{b}$ is the element-wise, or Hadamard, product between vectors
$\mbf{a}$ and $\mbf{b}$. $\mbf{x}^{1/2}$ is the element-wise
square-root of vector $\mbf{x}$. $\| \mbf{x} \|_{\mbf{W}} = \sqrt{ \mbf{x}^\top \mbf{W}  \mbf{x}}$ where $\mbf{W}$
is a positive semidefinite matrix. The modulus function is written
compactly as $\text{mod}_y(x) \triangleq \text{modulo}(x,y)$.

 \begin{figure}
 \centering
 \includegraphics[height=1.9in]{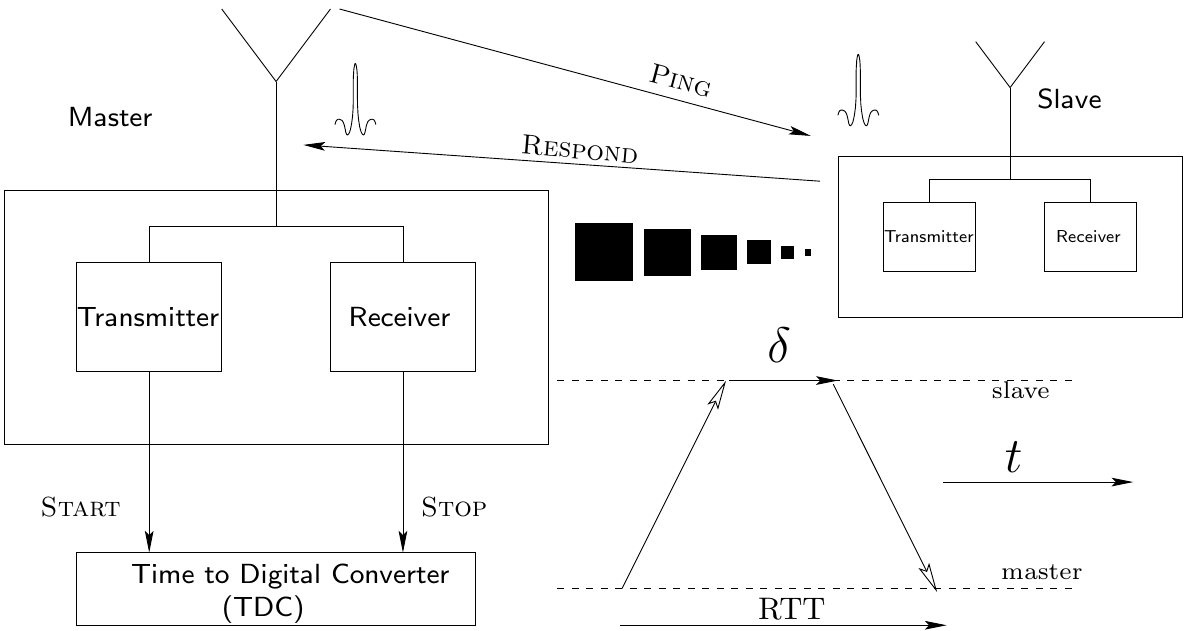}
 \caption{Principle of operation of the considered system: the master node sends a PING signal and the slave node responds. The master node is equipped with a time-to-digital converter capable of measuring the round-trip-time. }
 \label{fig:RTT}
 \end{figure}

\section{System Model and Problem Formulation} \label{sec:sys_model}

In this section, we first describe the principle of operation of the
considered system, then we illustrate the proposed clock measurement
mechanism. Subsequently, we define the related signal model and present a
basic estimator of the parameters of interest, based on existing literature.

\subsection{Principle of Operation: Round-Trip Time}

We consider two nodes equipped with a wireless transmitter and a
receiver. One being called the \emph{master} and the other called 
\emph{slave} \cite{DeAngelisEtAl2013}. We assume that the master node is capable
of measuring the time interval between the transmission of a \textsc{Ping} signal
and the reception of a \textsc{Respond} signal, sent by the slave, after a predetermined
delay. A conceptual overview is given in Fig.~\ref{fig:RTT}. 
This time interval is denoted as the \emph{round-trip-time}
(\textsc{RTT}) which we express as a function $\tilde{y}$ of
the unknown range between the master and slave $\rho$ and the delay $\delta$
\begin{equation}\label{eq:basicmodel}
\tilde{y} = \delta + \frac{2}{c}\rho,
\end{equation}
where $c$ is the speed of light in meters per second. Conventionally,
the delay is an ideal analog implementation and known at the master,
cf. \cite{schedule_comm_lett,DeAngelisEtAl2013}. Here, however, we
assume that all timed events are based on local clocks at each
node. As we will show below, this enables the joint estimation of both
range and parameters for clock synchronization.

Time is measured using a clock by counting a number of periods of the
clock signal
\begin{eqnarray*}
  C_0  &=& \frac{n}{f_0} + \varphi  
\end{eqnarray*}
where $n$ is the number of clock cycles, $f_0$ is the nominal clock frequency
and $\varphi \in [0, 1/f_0) $ is the initial phase of the clock when the measurement
begins. Any fractional deviation from the nominal clock frequency is
termed as skew of the clock, represented usually as $\alpha > 0$. Time is measured using a skewed clock, expressed in the nominal frequency $f_0$ as
\begin{eqnarray*}
  C  &=& \alpha\frac{n}{f_0} + \varphi,
\end{eqnarray*}
and the actual frequency $f$ deviates from nominal clock frequency given by $\alpha =  f_0 / f$. Equivalently the above model is
widely used as \cite{tretter_paper,skog_handel2,TDOA_clock_Gholami}:
\begin{eqnarray*}
  C  &=& \alpha t + \varphi  \, ,
\end{eqnarray*}

For simplicity and to establish the notation, let us first consider a
standard linear clock model of the observed times at master and slave nodes,
\begin{eqnarray*}
C_m  &=& \alpha_m t + \varphi_m 
\end{eqnarray*}
and
\begin{eqnarray*}
C_s  &=& \alpha_s t + \varphi_s,
\end{eqnarray*}
where $\alpha_m$ and $\alpha_s$ denote their clock skew with respect to nominal frequency $f_0$ and $\varphi$ is the initial phase offset.

\begin{figure} 
 \begin{center}
 \includegraphics[height=2.6in, width=2.6in]{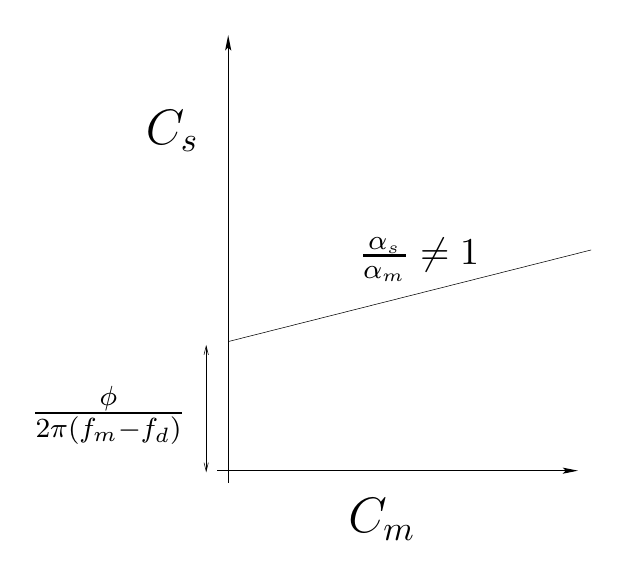}
 \caption{Relative clock error models, $(C_m,C_s)$ in
   \eqref{eq:rel_clk} $\alpha_s/\alpha_m$ denotes the clock skew
   and $\phi$ initial phase in radians.}
 \label{fig:CM}
\end{center} 
\end{figure}

The clock parameters are assumed time-invariant
\cite{Isaac_handel}. If $f_m$ and $f_s$ denote the clock frequencies
at the master and slave, then we define the frequency difference $f_d \triangleq f_m - f_s$. The relative
clock skews will then be parameterized by $f_d$ as $\alpha_s /
\alpha_m = f_m/f_s = f_m/(f_m - f_d)$. The time observed at the slave $C_s$ can then
be expressed relative to a reference, i.e. the master node clock $C_m$,
\begin{equation}\label{eq:rel_clk}
\begin{split}
C_s  &= \alpha_s \left(  \frac{C_m - \varphi_m}{\alpha_m} \right) +
\varphi_s\\
&= \frac{\alpha_s}{\alpha_m} C_m + \left( \varphi_s - \frac{\alpha_s}{\alpha_m}\varphi_m \right)
\\
&= \left( \frac{f_m}{f_m - f_d}\right) C_m + \frac{\phi}{2\pi (f_m-f_d)},
\end{split}
\end{equation}
where the second term is the relative clock offset in seconds and we correspondingly define the relative clock offset in radians $\phi$ as $$\phi \triangleq 2\pi (f_m-f_d) \left( \varphi_s - \frac{\alpha_s}{\alpha_m}\varphi_m \right) \in [0,2\pi).$$ This model is illustrated in
Fig.~\ref{fig:CM}. 
Given that $C_m$ is the
reference clock, $f_m$ is known and to perform synchronization of
master and slave clocks, only estimation of
$f_d$ and $\phi$ is necessary \cite{MSTINFOCOM}.

If we assume that the slave generates the delay by counting a fixed and known number
of clock cycles, equivalent to quantization in the time domain, the
nominal delay $\delta_0$ becomes dependent on the frequency difference and can be written as
\begin{equation*}
\begin{split}
\delta_0(f_d) & = \frac{\lfloor (C'_s - C_s ) (f_m - f_d) \rfloor}{ f_m - f_d} \, ,
\end{split}
\end{equation*}
where $C_s$ and $C'_s$ denote the time of reception of the \textsc{Ping} and transmission of the \textsc{Respond} signal at the slave. 

On the other
hand, the master is equipped with a TDC, and is therefore able to
measure time intervals with a much higher resolution than its clock
period. Based on this mechanism, which we will describe in more detail
below, the time measured at the master is given by an integer
number of slave clock cycles (i.e. $\delta_0$) plus a remainder
term, which is naturally modeled by a periodic modulus operation.
In the following subsection we present a model of this term, denoted
$h(t; f_d, \phi)$ which is dependent on the unknown clock parameters
$f_d$ and $\phi$. Therefore, we can write the delay as
\begin{equation}\label{eq:delaymodel}
\delta = \delta_0(f_d)  + h(t; f_d, \phi)
\end{equation}
and in combination with \eqref{eq:basicmodel} the round-trip-time
measurements will be informative of the range $\rho$ to the
slave but also of the relative clock parameters of the slave, $f_d$ and $\phi$, via $\delta$.

\subsection{Clock Measurement Mechanism}
\label{sec:mechanism}

\begin{figure*}
 \begin{center}
\hspace{-0.3in}
 \includegraphics[width=7.1in]{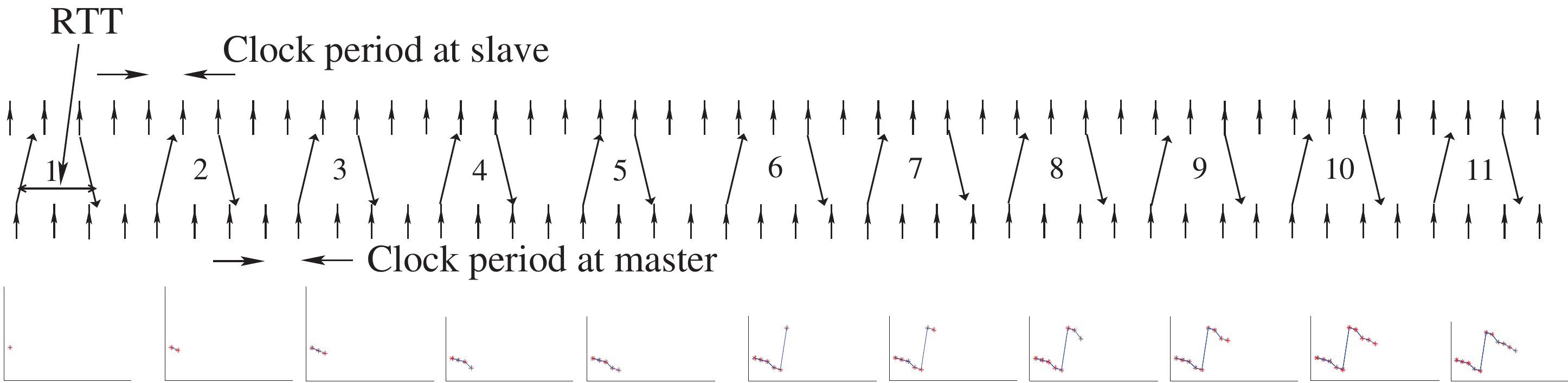}
 \caption{A diagram depicting the clock mechanism. Note that the RTT measurements, plotted below, vary depending on the shift of the slave clock edge relative to the master clock.}
 \label{fig:clock_mechanism}
 \end{center}
 \end{figure*}

Assuming the frequency difference is small, i.e. $f_d \ll f_m$, we can
neglect the dependence of $\delta_0$ on $f_d$ since $f_s = f_m - f_d
\simeq f_m$ and thus we can approximate $\delta_0$ as a known constant, i.e.
$$\delta_0 \simeq \frac{\lfloor (C'_s - C_s ) f_m \rfloor}{ f_m } \, .$$
Similarly, \eqref{eq:delaymodel} can be approximated at
a specific time instant $t$ as
\begin{equation}\label{eq:delaymodelmod}
\begin{split}
\delta &= \frac{\lfloor (C'_s - C_s) (f_m-f_d) \rfloor}{f_m - f_d} + \frac{1}{2 \pi
  (f_m-f_d)} \text{mod}_{2 \pi} (f_d t + \phi)\\
&\simeq \delta_0 + \frac{1}{2 \pi
  f_m} \text{mod}_{2 \pi} (f_d t + \phi) 
\end{split}
\end{equation}
where periodic modulus operation represents the remainder and is
dependent on the clock offset and frequency difference. We now
illustrate how the remainder term appears at the master node. 
Consider the following example based on the events described in Fig.~\ref{fig:RTT}:
 \begin{figure}
 \centering
 \includegraphics[height=3in]{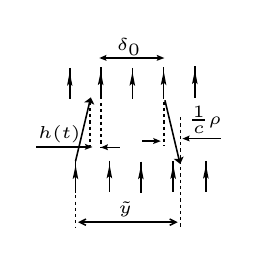}
 \caption{A space-time diagram depicting the clock edges (vertical
   arrows) at the slave (top row) and master (bottom row). The diagram
   relates the quantities in \eqref{eq:basicmodel} and
   \eqref{eq:delaymodel}.  Specifically, the remainder term $h(t)$ in \eqref{eq:delaymodel}
   is highlighted.}
 \label{fig:remainder}
 \end{figure}
\begin{enumerate}
\item Master sends a \textsc{Ping} pulse on a positive edge of clock, TDC receives the
  \textsc{Start} signal instantly and starts measuring time.
\item On receiving the pulse, slave starts the delay generation
  from subsequent positive edge of its own clock. 
\item Slave sends \textsc{Response} after the delay $\delta_0$ (say, two
  clock cycles in this example)
\item Master receives the \textsc{Response}, \textsc{TDC} gets the
  \textsc{Stop} signal and measures the \textsc{RTT}.
\end{enumerate}
This sequence of events is also illustrated in the space-time diagram of
Fig.~\ref{fig:remainder} where the remainder term $h(t)$ is
highlighted. The master repeats the above transmission procedure with a fixed
periodicity. The result is illustrated in Fig.~\ref{fig:clock_mechanism} which shows the role of the clock in
repeated pings.
At instance \emph{`5'} \textsc{Ping} from master
  is received by slave at a clock edge and hence results in smallest
  RTT. Whereas, at instance \emph{`6'} slave receives \textsc{Ping}
  narrowly after the clock edge and hence waits nearly a full clock duration to
  start its delay count of a clock cycle. This results in largest RTT
  measurement at master. As can be seen, the clock offset and frequency
 results in a perodic time-varying remainder term $h(t)$ with a sawtooth-like waveform
 corresponding to the modulus function that follows from rounding.

In addition to the effect of the rounding operation, the remainder
term in \eqref{eq:delaymodelmod} is also subject to zero-mean random clock
jitter, which we denote $v$. At measurement time instant $t_i$ we
therefore model the remainder as
\begin{equation}\label{eq:remainderterm}
h(t_i;f_d, \phi) = \frac{T_m}{2\pi} \text{mod}_{2 \pi}(f_d t + \phi + v(t_i)),
\end{equation}
where $T_m = 1/f_m$. Finally, inserting \eqref{eq:remainderterm} into \eqref{eq:delaymodel} under
$f_d \ll f_m$, \eqref{eq:basicmodel} results in the following the
RTT measurement model at time $t_i$:
\begin{equation}\label{eq:model_scalarreduced}
y(t_i) = \frac{T_m}{2 \pi} \text{mod}_{2 \pi} ( f_d t_i  + \phi + v(t_i)
) + \delta_0 + \frac{2}{c} \rho + n(t_i),
\end{equation}
where $n(t_i)$ denotes zero-mean measurement noise. Figure
\ref{fig:sawtooth} shows a measurement capture of above phenomena on
our flexible UWB radio test-bed \cite{DeAngelisEtAl2013}. In Section~\ref{sec:residualanalysis} we validate the
model \eqref{eq:model_scalarreduced} by means of residual analysis,
justifying the approximations used above for the considered experimental
setup.

\subsection{Signal Model and Measurements}

Let $\mbf{y} = [y(t_1) \: \cdots \: y(t_N)]^\top \in \mathbb{R}^N$ and $\mbf{t} = [t_1
\cdots \: t_N]^\top \in \mathbb{R}^N$ be the vector of $N$
measurements and time instances, respectively. Then model
\eqref{eq:model_scalarreduced} can be written compactly in vector form:
\begin{equation}\label{eq:model_vectorreduced}
\mbf{y} = \mbf{h}(f_d,\phi,\mbf{v}) + \delta_0 \mbf{1} + \frac{2\rho}{c}
\mbf{1} + \mbf{n} \, ,
\end{equation}
where 
\begin{equation*}
\mbf{h}(f_d,\phi, \mbf{v}) \triangleq \frac{T_m}{2 \pi} \text{mod}_{2
  \pi}(f_d\mbf{t} + \phi \mbf{1} + \mbf{v} ) \, . 
\end{equation*}
The vectors $\mbf{v}$ and
$\mbf{n}$ contain the random noise terms with unknown variances $\sigma^2_v$ and $\sigma^2_n$,
respectively. The goal is to estimate $\rho$, $f_d$ and $\phi$ from
$\mbf{y}$.

Central to \textsc{RTT} measurement
procedures is the device called time-to-digital converter (TDC) as is shown
in Fig.~\ref{fig:RTT}. Commercially available TDCs can
measure time with resolution up to $50$\,ps \cite{acam_TDC}. As shown in
Fig.~\ref{fig:RTT}, TDC measures the time difference between two events and
thus measures \textsc{RTT}.  The TDC has its own clock for coarse
counting \cite{acam_TDC}.

\begin{figure}
\begin{center}
\includegraphics[width=3.5in]{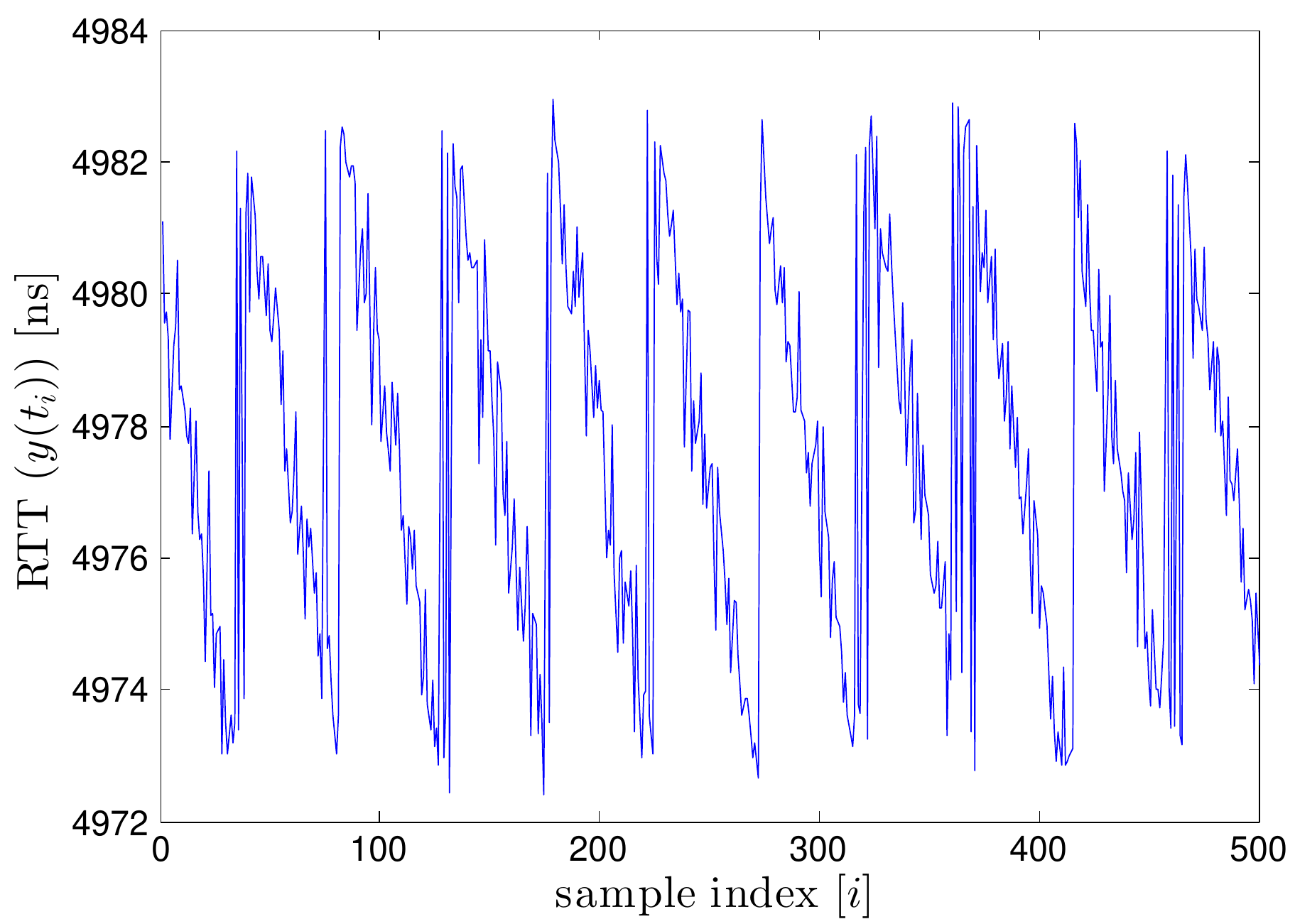}
\caption{Experimental data record showing the ``sawtooth'' behavior of the
  \textsc{RTT} measurements. The data was obtained with the master and slave
  antennas placed 0.5 m apart, $\delta_0 = 4900$\,ns.}
\label{fig:sawtooth}
\end{center}
\end{figure}

Here we wish to highlight a few important features:
\begin{itemize}
\item Measurement model \eqref{eq:model_vectorreduced} arises from sampling the slave
  clock over a wireless medium and measuring the
  relative clock drift using a TDC at master node.
\item No timestamping and hence no data transfer is required between master and
  slave devices.
\item Traditionally, clock parameters are estimated as shown in Fig.~\ref{fig:CM},
  where measured time monotonically increases. Long time measurements using
  a clock require high accuracy. Whereas, RTT measurements as seen in
  Fig.~\ref{fig:sawtooth} require
  short time measurements, hence clock accuracy and stability is a
  lesser concern. When measuring smaller time duration the TDC clock
  skew is assumed to be negligible.  
\item Clock parameter information is available in the time domain as well as the frequency domain. The latter fact facilitates extracting information using frequency-based methods. 
\end{itemize}
Next, we introduce a conventional estimator for estimating $f_d$,
$\phi$ and $\rho$
by unwrapping the saw-tooth measurement \eqref{eq:model_vectorreduced}; an early work is
\cite{tretter_paper}, cf. discussion in \cite{volker2002frequency,Isaac_handel}.

\subsection{Unwrapped Least Squares (ULS)}
We consider the model in \eqref{eq:model_scalarreduced} and formulate
a three-step unwrapped least-squares method (ULS). First, by
approximating the arithmetic average of the modulus term in
\eqref{eq:model_scalarreduced} as the amplitude of the sawtooth-like
waveform, $T_m/2$, we obtain 
\begin{align*}
\bar{y} &= \frac{1}{N} \sum^{N}_{i=1} y(t_i) \\
&\simeq \frac{T_m}{2} + \delta_0 + \frac{2\rho}{c}+ \frac{1}{N} \sum^{N}_{i=1} n(t_i).
\end{align*}
From this we form an approximate least-square estimate of the range
\begin{align*}
\hat{\rho} &= \argmin_{\rho} \; \left( \bar{y} - \frac{T_m}{2} -
  \delta_0 - \frac{2\rho}{c}\right)^2 \\
&= \frac{c}{2}
\left( \bar{y}   - \frac{T_m}{2} - \delta_0 \right) \,.
\end{align*}
Then, we normalize the measurements $y(t_i)$ to the $\left[ -\pi, \pi \right)$-interval as follows:
\begin{align*}
z(t_i) \triangleq \frac{2\pi}{T_m}
\left( y(t_i)  - \bar{y} \right) \,. 
\end{align*}
Next, we apply a `phase-unwrapping' algorithm, denoted by $\mathcal{U}(\mbf{z})$ where
$\mbf{z} = [z(t_1) \cdots z(t_N)]^\top$ \cite{Isaac_handel,tretter_paper,matlab_unwrap}. This algorithm adds multiples of $\pm 2 \pi$ rad to the phase when the absolute difference between consecutive elements is greater than $\pi$ rad.
Finally, we obtain $\hat\phi$ and  $\hat f_d$ 
by solving the following linear least-squares problem on the unwrapped samples:
\begin{equation}
\begin{split}
[\hat{f}_d, \hat{\phi}]&= \argmin_{f_d, \: \phi } \: \left \| \mathcal{U}(\mbf{z})
- f_d \mbf{t} - \phi \mbf{1} \right \|^2_2.
\end{split}
\end{equation}
Whilst the ULS estimator is computationally efficient, erroneously
unwrapped samples by $\mathcal{U}(\cdot)$ which occur in poor signal conditions lead to performance degradation. 

Furthermore, as formulated it is not robust with respect to noise outliers and its estimates may deteriorate significantly.
The presence of outliers is a common issue in practical wireless measurement systems. As an evaluation example, the wireless localization scenario is considered in \cite{ZoubirEtAl2012}. In such a scenario, outliers in the time-of-arrival measurements are present due to propagation channel conditions, and robust regression techniques are employed for the purpose of reducing the adverse impact of outliers on localization performance.

\section{Proposed Estimators}
\label{sec:estim}

In view of the limitations of the unwrapped least squares approach
described above, we develop two alternative methods to estimate the
offset frequency $f_d$, offset phase $\phi$ and range $\rho$ 
in \eqref{eq:model_vectorreduced}.

\subsection{Periodogram and Correlation Peaks (PCP)}

In the first approach we exploit the periodicity of the signal model
\eqref{eq:model_scalarreduced} to obtain a frequency estimate. Using the
periodogram \cite{Stoica&Moses2005_spectral} we can compute an
estimate up to an unknown sign,
\begin{equation*}
\check{f}_d = \argmax_{f_d \in F} \; \left| \sum^N_{i=1} y(t_i) e^{- j 2\pi f_d t_i} \right|^2,
\end{equation*}
where $F \subset [0, f_{\text{max}}]$ is a uniform grid of frequencies.
Under uniform sampling, the estimate can be computed efficiently using
the fast Fourier transform. Note that $f_d$ determines the slope of
the periodic sawtooth waveform in \eqref{eq:model_scalarreduced}, cf. Fig.~\ref{fig:sawtooth}. The sign of
$f_d$ can be resolved jointly with phase estimation by correlating the
observed signal with sawtooth waveforms based on $\check{f}_d$ and
$-\check{f}_d$, corresponding to positive
and negative slopes.

Then using the estimate $\check{f}_d$ a waveform
$\mbf{p}(\phi,s) = \text{mod}_{2  \pi}(s\check{f}_d\mbf{t} + \phi
\mbf{1})$ is generated, for a fixed frequency sign $s = \{ -1, 1\}$
and nominal
phase $\phi$. The phase estimate, and the resolved sign of the frequency, is
then obtained by the correlation peak,
\begin{equation*}
[\hat{\phi}, \hat{s}] = \argmax_{\phi \in \Phi, \: s \in \{-1,1\}} \; |\mbf{y}^\top \mbf{p}(\phi,s)|,
\end{equation*}
where $\Phi \subset [0, 2\pi)$ is a uniform grid of phases. The frequency estimate is $\hat{f}_d = \hat{s}\check{f}_d$.
Finally, the range estimate is obtained by a direct least-squares fit
using \eqref{eq:model_vectorreduced} which results in a particularly simple
form. Using vector notation, it can be written as
\begin{align*}
\hat{\rho} &= \argmin_{\rho} \left\| \mbf{y} -
\frac{T_m}{2\pi}\mbf{p}(\hat{\phi},\hat{s}) - \delta_0 \mbf{1} - \frac{2\rho}{c}\mbf{1} \right\|^2_2\\
&=\frac{c}{2N}\mbf{1}^\top \left(\mbf{y} -
  \frac{T_m}{2\pi}\mbf{p}(\hat{\phi},\hat{s}) - \delta_0 \mbf{1} \right).
\end{align*}

We summarize this low-complexity three-step approach as computing the
periodogram and correlation peaks (PCP). This approach circumvents the
need for phase unwrapping the data and utilizes frequency domain
characteristics of measurements.

%
%
%
%
%




\subsection{Weighted Least Squares (WLS)} \label{s:WLS}

In the second approach we address the issue of robust estimation in
the presence of noise outliers, using a weighted squared-error criterion
\begin{equation}
J(f_d, \phi, \rho) \triangleq \left \| \mbf{y} -
  \mbf{h}(f_d,\phi, \mbf{0}) - \delta_0 \mbf{1}  - 2c^{-1} \rho \mbf{1} \right \|^2_\mbf{W},
\label{eq:J_wls}
\end{equation}
and find the minimizing arguments. Here $\mbf{W} = \text{diag}(\mbf{w}) \succeq \mbf{0} \in
\mathbb{R}^{N \times N}_+$ is a diagonal weight
matrix. The weights $\mbf{w}$ can be chosen to mitigate the effect of
outliers by downweighting the samples likely to be outliers, as discussed in the next subsection. This makes WLS robust with respect to outliers. Setting uniform weights $\mbf{w} \propto \mbf{1}$ reduces \eqref{eq:J_wls} to a standard least-squares criterion.

We begin by minimizing \eqref{eq:J_wls} with respect to $\rho$. This
gives
\begin{equation}
\label{e:WLS_rho}
\hat{\rho} = \frac{c \mbf{w}^\top (\mbf{y} - \mbf{h}(f_d,\phi,\mbf{0})
  -  \delta_0\mbf{1})}{2
  \mbf{1}^\top \mbf{w}} \;.
\end{equation}
Inserting the estimate \eqref{e:WLS_rho} back into \eqref{eq:J_wls} and defining
$\mbf{r}(f_d,\phi) \triangleq \mbf{y} - \mbf{h}(f_d,\phi, \mbf{0}) -
\delta_0 \mbf{1}$, we can write the minimization with respect to $f_d$
and $\phi$ as a two-dimensional grid search 
\begin{equation}
\label{e:WLS_estimator}
\begin{split}
[\hat{f}_d, \hat{\phi}]&= \argmin_{f_d \in F, \: \phi \in \Phi} \: \| \mbf{w}^{1/2} \odot \mbf{r}(f_d, \phi) \|^2_2 - \frac{| \mbf{w}^\top \mbf{r}(f_d, \phi) |^2}{\mbf{1}^\top \mbf{w}},
\end{split}
\end{equation}
where $F \subset
[-f_{\text{max}}, f_{\text{max}}]$ and $\Phi \subset
[0, 2\pi)$ denote the grids. See the appendix for the derivation.

\subsection{Choosing the Weights for Outlier-prone Data}



For applying the WLS method derived in Section \ref{s:WLS} to outlier prone data, we propose to choose the $i$-th element $w_i$ of the weight vector $\mbf{w}$ as follows:
\begin{align} \label{e:outl_criterion}
w_i = \left\{
\begin{array}{cl}
	1 & \mbox{if } \left| y(t_i) - \hat{\mu}_{1 / 2}\left(\mbf{y}\right) \right| \leq 3 \hat{\sigma}_{\text{mad}}\left(\mbf{y}\right) \\
	0 & \mbox{otherwise,}
\end{array}
\right.
\end{align}
where $\hat{\mu}_{1 / 2}\left( \cdot \right)$ denotes the sample median operator, and $\hat{\sigma}_{\text{mad}}$ denotes the normalized Median Absolute Deviation (nMAD), which is defined, cf. \cite{ZoubirEtAl2012}, as
\begin{align*}
\hat{\sigma}_{\text{mad}}\left( \mbf{y} \right) = 1.483  \cdot
\hat{\mu}_{1 / 2}\left(\mbf{y} - \mbf{1} \hat{\mu}_{1 / 2}\left(\mbf{y}
  \right) \right) \,.
\end{align*}
As the signal model \eqref{eq:model_scalarreduced} has no trend and a constant offset, this choice of weights \eqref{e:outl_criterion} takes large deviations from the median to be an indicator of a likely outlier and downweights the corresponding samples.

We note that other outlier-rejection strategies are possible \cite{rousseeuw1984least}, but are not presented here; see \cite{Hodge&Austin2004} for an extensive survey. An advantage of the proposed strategy is that it is automatic, since it does not require any input or parameter tuning by the user. Moreover, in the considered context, another desirable feature of the strategy is its tight integration in the WLS framework. Finally, the proposed strategy gives good results in a practical outlier-prone scenario, as we show in the experimental evaluation of Section \ref{sec:experiments}.
%




%

\section{Numerical Results} \label{sec:numerical}
In this section, we explore the properties of the model in \eqref{eq:model_vectorreduced} and evaluate the performance of the proposed estimators in realistic scenarios. 

\subsection{Simulation Setup}
The numerical simulations have been performed, unless otherwise indicated, using the following fixed values of the parameters of interest: $f_d = -32$\,Hz, $\rho = 2$\,m. For each Monte Carlo iteration, the initial phase $\phi$ is set to a random value according to a uniform distribution in the $\left[ 0, 2 \pi \right)$ interval. 
Unless otherwise noted, a record comprised of $N = 100$ samples is
generated for each Monte Carlo iteration according to the model in
\eqref{eq:model_vectorreduced}. The true value of the master clock frequency is
$T_m=10^{-8}$\,s, the value of $\delta_0$ is $5\times 10^{-6}$\,s and a measurement update period of $T_s = 10^{-3}$\,s has been used. 
Taking into account the model in \eqref{eq:model_vectorreduced}, we define 
$$\mbox{SNR}_c \triangleq 10 \log_{10}\frac{T^2_m}{\sigma_n^2} \,, \quad
\mbox{SNR}_j \triangleq  10 \log_{10}\frac{\left(2\pi\right)^2}{\sigma_v^2}\,.$$
$\mbox{SNR}_c$ is signal-to-noise ratio over the wireless channel and
$\mbox{SNR}_j$ is signal to noise ratio of the clock jitter noise
\cite{jitter_ref}. The channel noise for simulations is generated in seconds and the jitter
noise is the dimensionless phase noise as can be seen from \eqref{eq:model_vectorreduced}. 
The performances are evaluated using the root mean-square error (RMSE)
and computed by averaging over 1000 Monte Carlo iterations.

\begin{figure*}
\centering
\subfigure
  {\includegraphics[]{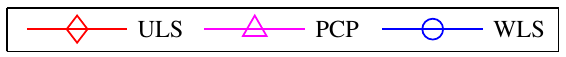}}
\\  
\vskip -0.3cm
\setcounter{subfigure}{0}
\subfigure[]
  {\includegraphics[width = 0.32  \textwidth]{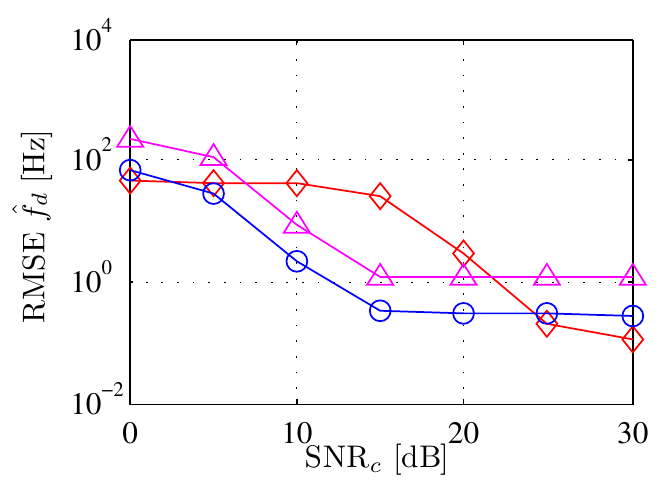}}
\subfigure[]  
  {\includegraphics[width = 0.32  \textwidth]{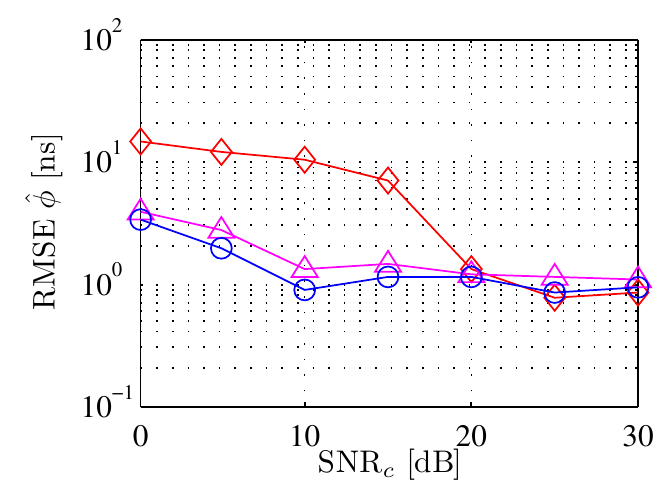}}
\subfigure[]  
  {\includegraphics[width = 0.32  \textwidth]{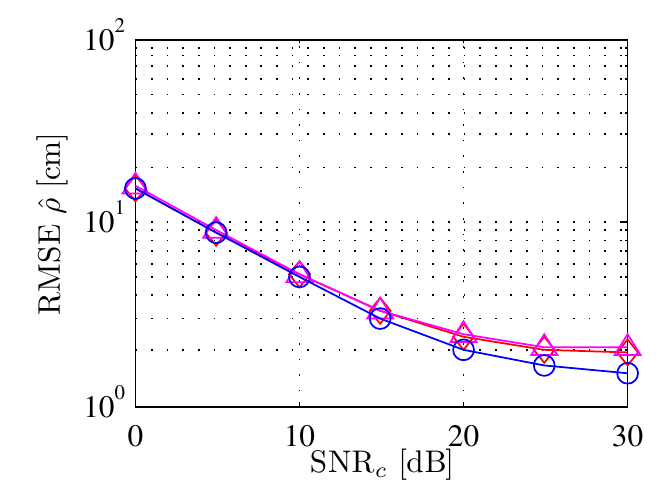}}  
\\
\vskip -0.3cm
\subfigure[]
  {\includegraphics[width = 0.32  \textwidth]{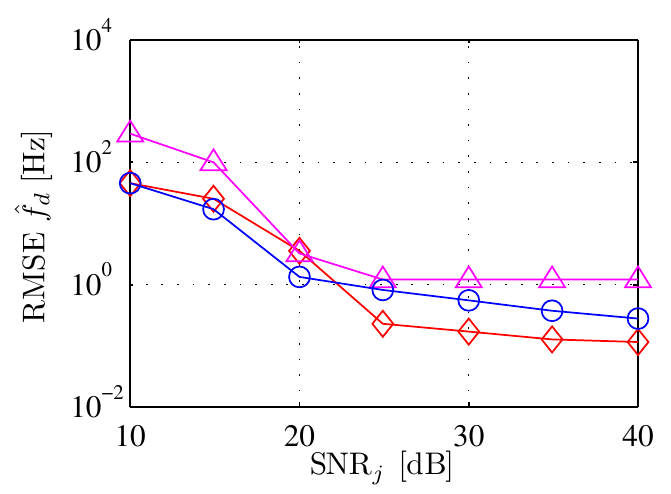}}
\subfigure[]  
  {\includegraphics[width = 0.32  \textwidth]{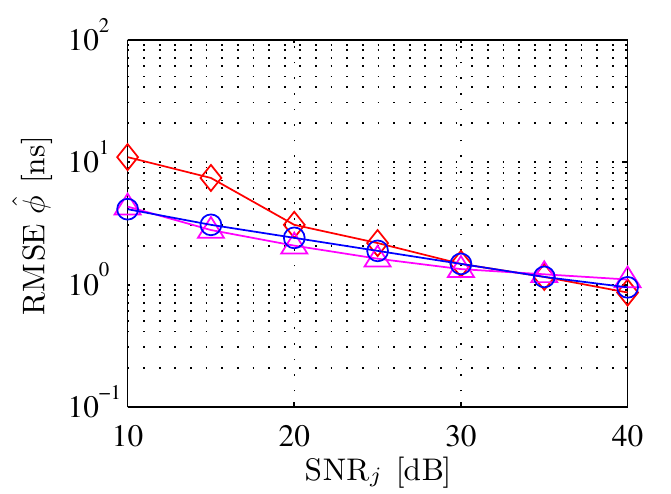}}
\subfigure[]  
  {\includegraphics[width = 0.32  \textwidth]{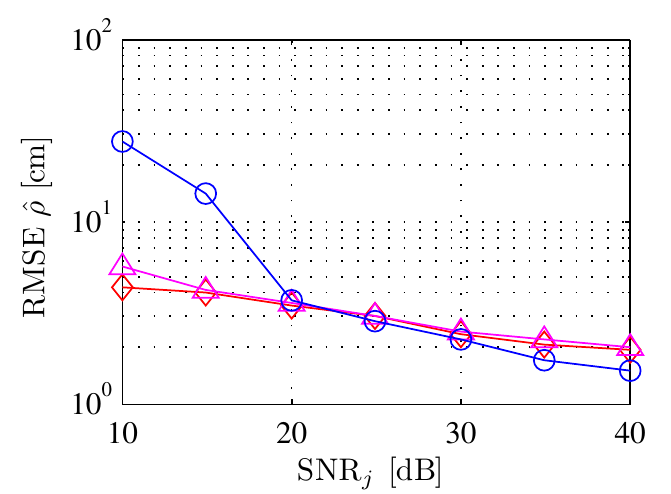}}  
\\
\vskip -0.3cm
\subfigure[]
  {\includegraphics[width = 0.32  \textwidth]{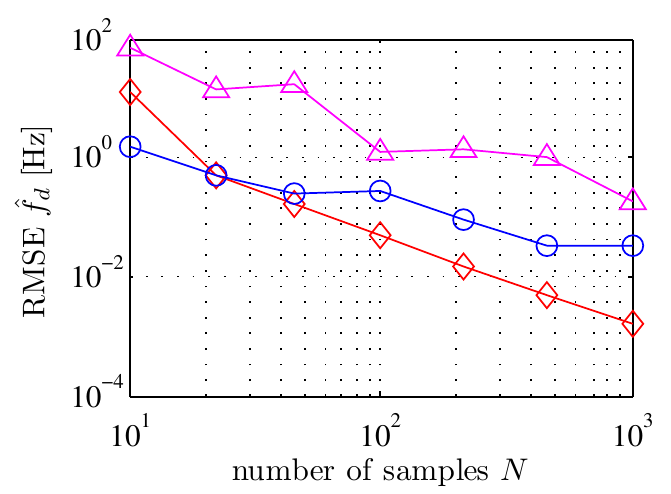}}
\subfigure[]  
  {\includegraphics[width = 0.32  \textwidth]{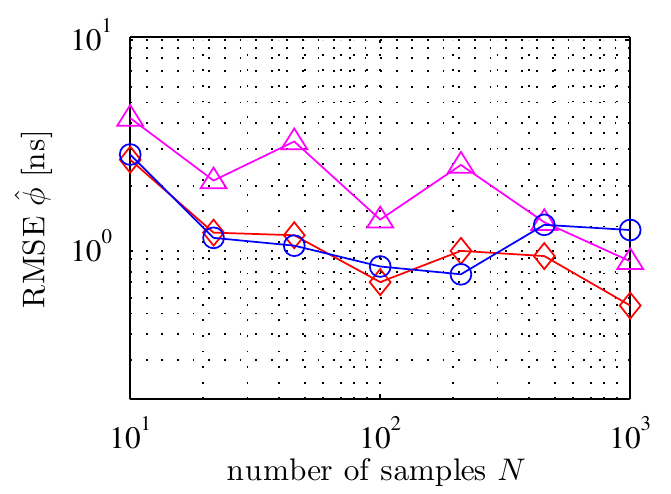}}
\subfigure[]  
  {\includegraphics[width = 0.32  \textwidth]{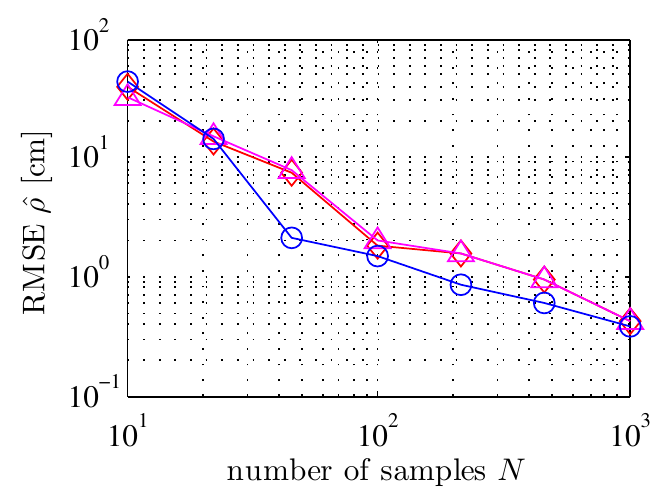}}    
\\
\vskip -0.3cm
\subfigure[]
  {\includegraphics[width = 0.32  \textwidth]{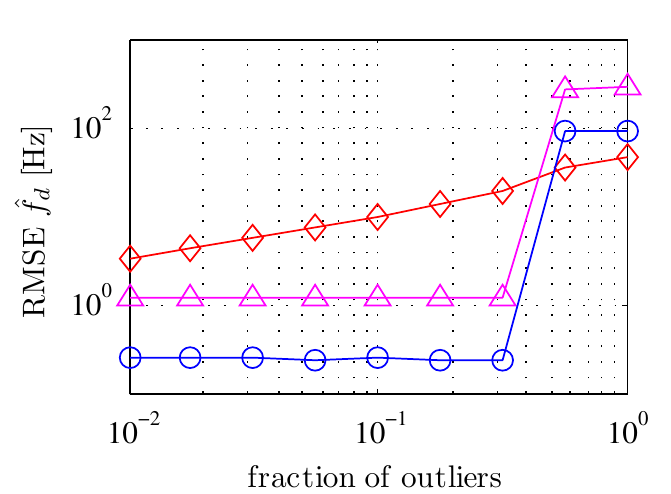}}
\subfigure[]  
  {\includegraphics[width = 0.32  \textwidth]{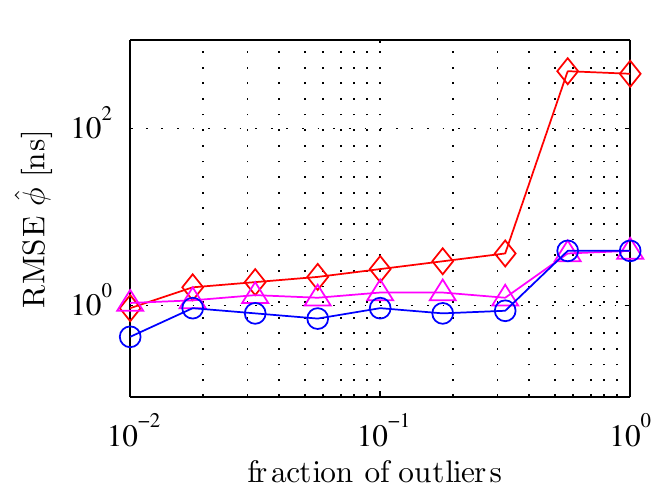}}
\subfigure[]  
  {\includegraphics[width = 0.32  \textwidth]{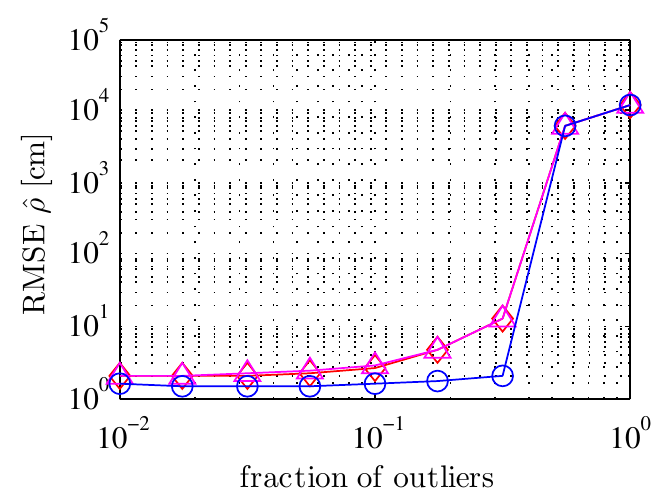}}    
\caption{Simulation results, averaged over 1000 Monte Carlo
  iterations. (a)-(c) varying SNR$_c$, with SNR$_j = 40$\,dB. (d)-(f)
  varying SNR$_j$, with SNR$_c = 30$\,dB. (g)-(i) varying the number of
  samples $N$, with SNR$_j = $ SNR$_c = 40$\,dB. (j)-(l) in the presence
  of outliers, varying the fraction of randomly-inserted outliers,
  with SNR$_j = $ SNR$_c = 40$\,dB. }
\label{f:simulations}
\end{figure*}

\subsection{Simulation Results}
A complete characterization of the Monte Carlo simulation results is provided in Fig. \ref{f:simulations}. 
Due to the fixed resolution of the search grid, WLS and PCP exhibit quantization effects, as shown e.g. in the asymptotic constant behavior at high SNR in Fig. \ref{f:simulations}\,a. Potentially, these effects can be overcome by refining estimates using standard interpolation, zero padding, or adaptive grid search methods.
However, their performance is still satisfactory. In fact, all considered estimators are able to obtain accuracy of the order of $1$\,Hz or better, when SNR$_c$ and SNR$_j$ are greater than $20$\,dB. 

Furthermore, we studied the behavior of the proposed methods as the
number of samples $N$ varies. The simulation results are shown in
Fig. \ref{f:simulations}\,g - \ref{f:simulations}\,i. The proposed WLS estimator shows improved accuracy in small-sample conditions with respect to ULS and PCP, which is advantageous in scenarios with slowly time-varying parameters. 
The performance of ULS at high SNR and for large $N$ is generally better than PCP and WLS, as shown in Fig. \ref{f:simulations}\,a, d, and g. 
This is consistent with the expected behavior. As noise vanishes, in fact, the unwrapping procedure results in a linear trend, which can be estimated statistically efficiently by the linear least-squares method.

Finally, in order to analyze the effect of outliers, we performed a set of Monte Carlo simulations substituting a fraction of the data with outliers, uniformly distributed in the $\left[ 3.5\cdot 10^{-6} \,, 4.9\cdot10^{-6} \right]$\,s interval.
The outliers are inserted at randomly-selected locations, yielding the
results in Fig. \ref{f:simulations}\,j - \ref{f:simulations}\,l. The
proposed WLS estimator has considerably improved robustness with respect to the
other considered methods. Moreover, it exhibits a threshold behavior,
reaching a breakdown point when the fraction of outliers is
approximately equal to $40\%$. On the other hand, ULS shows a degraded
performance even at a low fraction of outliers, in
Fig. \ref{f:simulations}\,g and \ref{f:simulations}\,h, therefore
validating the proposed WLS approach. PCP operates first in the
frequency domain which gives it a degree of insensitivity to outliers.

\subsection{Figure of Merit} 
The numerical results in Fig. \ref{f:simulations} show that, using the
considered RTT clock measurement mechanism and the proposed WLS
estimator, 100 samples are sufficient to achieve an estimation
accuracy of Hertz-order for 
clock frequency difference, nano-second order for initial phase
difference  and decimeter-order for ranging in practical noisy and outlier-prone conditions. This can be verified from the curves in Fig. \ref{f:simulations}\,(g)-(i), where an SNR$_c = $SNR$_j = 40$\,dB is considered. Then, from Fig. \ref{f:simulations}\,(a)-(f) it is possible to observe that such accuracy level is still achieved for SNR$_c$ as low as $10$\,dB and SNR$_j$ as low as $20$\,dB. More importantly, this accuracy is achieved by the proposed WLS estimator when up to 30$\%$ of the measured data are outliers, as shown in Fig. \ref{f:simulations}\,(j)-(l). 
Notice that $N=100$ samples correspond to an observation time of $0.1$\,s for the considered update period of $T_s = 10^{-3}$\,s, which we have chosen to be of the same order of magnitude as our practical experimental setup. Such an observation time is feasible for most practical applications of joint wireless synchronization and ranging.

\begin{figure*}
\centering
\subfigure
  {\includegraphics[]{legend.pdf}}
\\  
\vskip -0.3cm
\setcounter{subfigure}{0}
\subfigure[]
  {\includegraphics[width = 0.32  \textwidth]{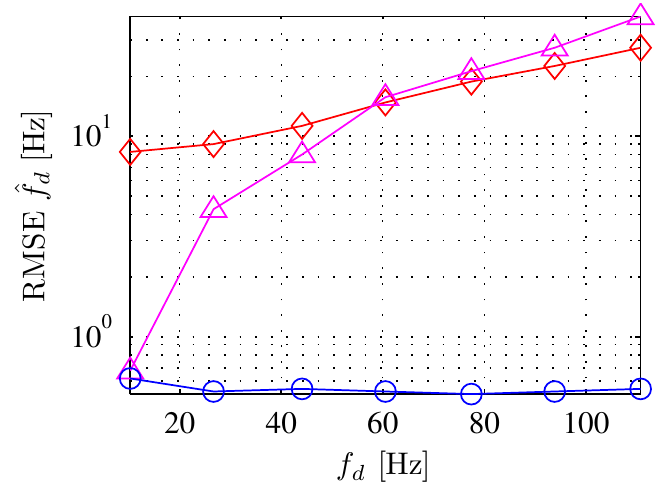}}
\subfigure[]  
  {\includegraphics[width = 0.32  \textwidth]{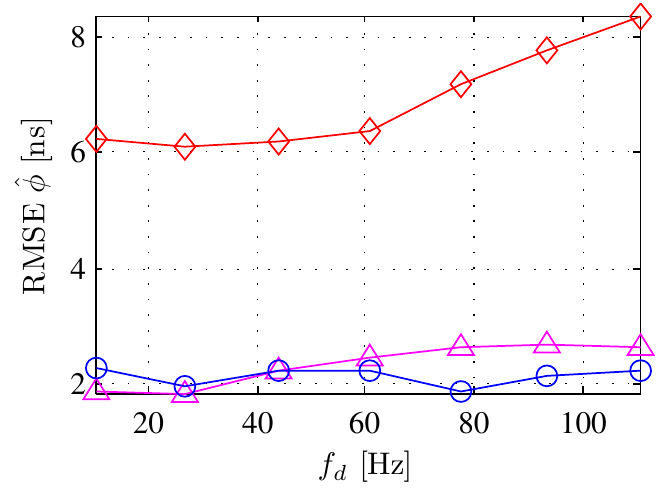}}
\subfigure[]  
  {\includegraphics[width = 0.32  \textwidth]{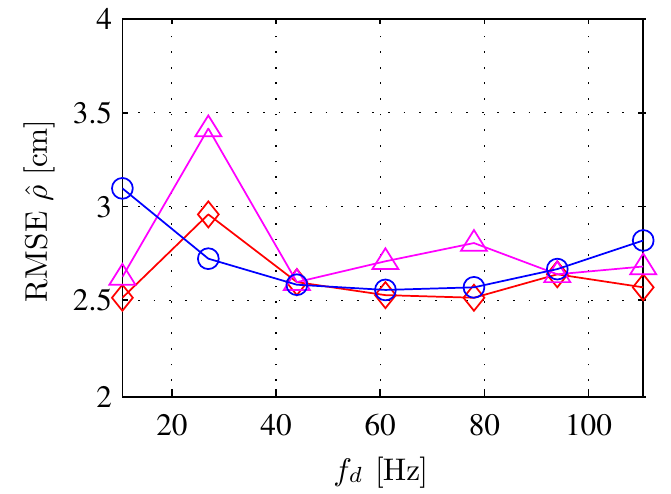}}  
\caption{Numerical simulations: RMSE performance of the proposed estimators for different values of the parameter $f_d$. Here, SNR$_c = 20$\,dB, SNR$_j = 20$\,dB, $N = 200$, and no outliers are present. Results are obtained by averaging over 1000 Monte Carlo iterations.}
\label{f:simulations_f_d}
\end{figure*}

\subsection{Sensitivity to Parameter Variations} 
To evaluate the performance of the considered estimators when the parameter $f_d$ varies, we provide the simulation results in Fig. \ref{f:simulations_f_d}, showing that WLS is insensitive to the value of $f_d$. On the other hand, ULS and PCP show considerable degradation of the frequency difference estimation performance as $f_d$ increases, see Fig. \ref{f:simulations_f_d}(a). 
As far as the range is concerned, we note that the model in \eqref{eq:model_vectorreduced} is linear in the range $\rho$, thus it is expected from the theory that the behavior does not change significantly for different values of $\rho$. This is confirmed numerically in Fig. \ref{f:simulations_f_d}(c), where the observed performance is approximately constant.

\section{Experimental Results} \label{sec:experiments}

We applied the proposed estimators to experimental data obtained from
an in-house test-bed which is based on the IR-UWB technology \cite{DeAngelisEtAl2013}.
In the following, we provide a description of the experimental setup,
we evaluate the performance of the considered estimators (ULS, PCP, WLS) on measured data, and eventually
we discuss the validation of the proposed model in
\eqref{eq:model_vectorreduced}. 
\subsection{Experimental Setup}

\begin{figure} 
 \centering
 \includegraphics[width = 0.98 \columnwidth]{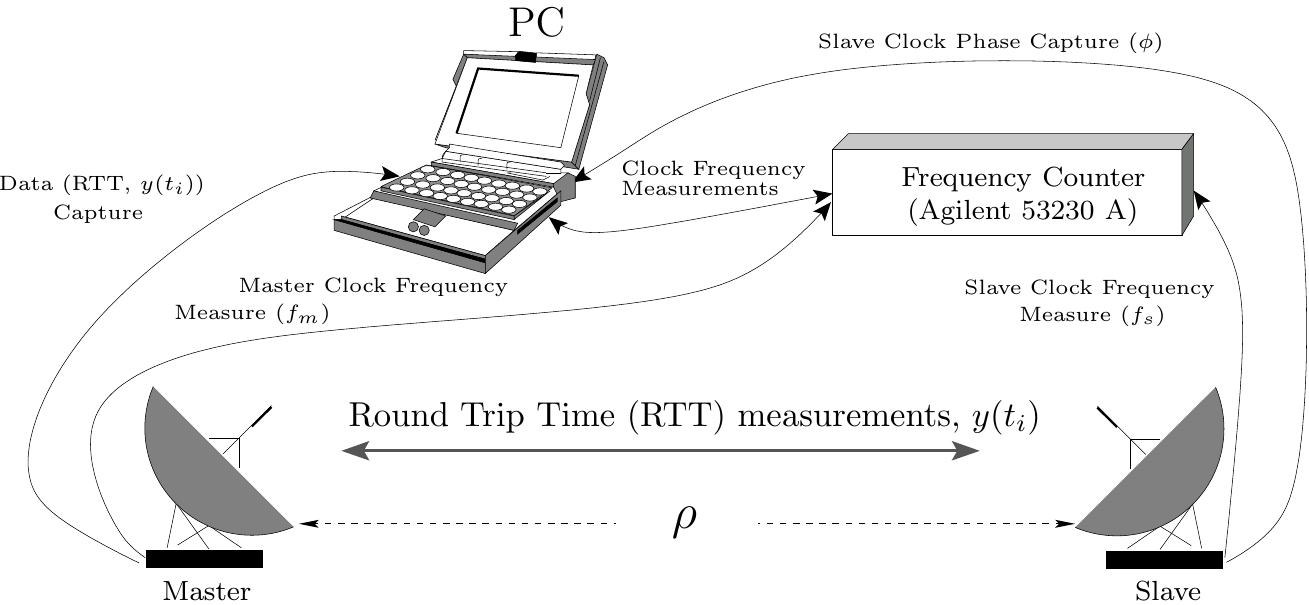}
 \caption{Diagram of the experimental setup.}
 \label{f:experimental_setup}
 \end{figure}

Experimental setup is shown in Fig.~\ref{f:experimental_setup}. Each of the two nodes was running on its own clock which was provided by a local oscillator in the ML505 evaluation board using a Xilinx Virtex-5 FPGA \cite{Xilinx_ML505}. Since the two local oscillators were not synchronized, a non-zero frequency difference between the two oscillators was present.
In the experimental setup, the master node was programmed to repeat the
\textsc{RTT} measurement procedure with a fixed sampling rate of $5$\,kHz,
and then transfer each result from the TDC to the host computer, where it
was stored and used for offline processing. The apparatus and components used in
  the experiments are relatively low cost and easily available. This
  brings the proposed idea closer to practice. 

As can be seen from
Fig. \ref{f:experimental_setup}, while capturing measurements, we capture
ground-truth values of the parameters. The ground-truth values are used for analyzing RMSE of the
estimates over measured data. The ground truth of range ($\rho$) is measured using a
handheld laser distance meter \cite{laser_ref}. Ground-truth values of 
master and slave clock frequencies ($f_m, f_s$) are
captured using Agilent 53230A frequency counter
\cite{agilent_counter}. Ground-truth of phase ($\phi$) is obtained by
capturing the clock state using primitive delays in FPGA. The clock of
slave node is propagated through these delays and the received first \textsc{Ping}
in a sequence \textsc{Ping}s from master triggers capture of the clock
phase \cite{Satyam_ISPCS}.  

\begin{figure}
 \centering
 \includegraphics[width = 0.98 \columnwidth]{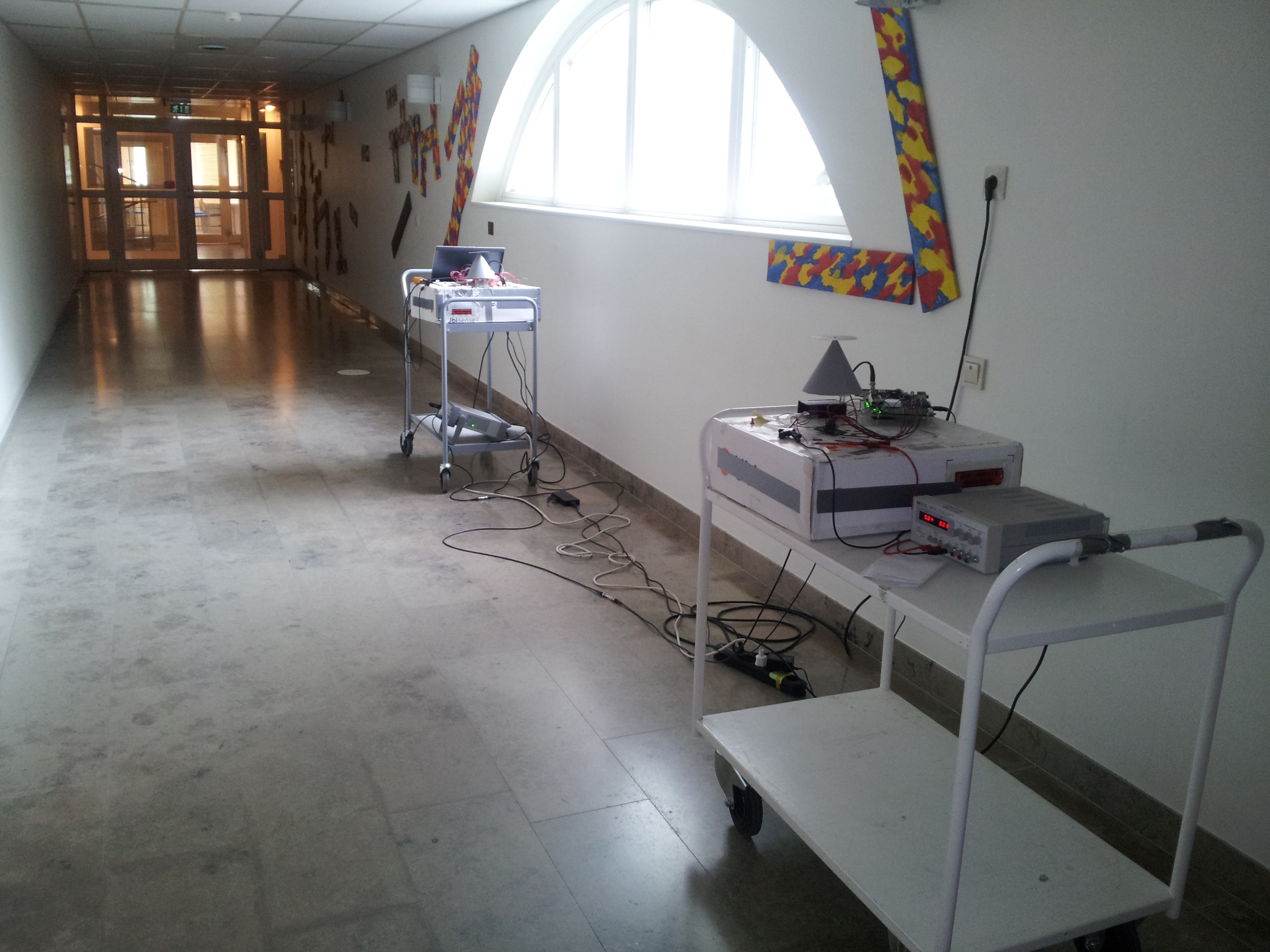}
 \caption{Picture of the experimental setup.}
 \label{f:photo_experiments}
 \end{figure}

We placed the master and slave nodes on two carts in line-of-sight
conditions in the office-like indoor environment depicted in
Fig. \ref{f:photo_experiments}. For each considered distance a record of 65356 \textsc{RTT} measurement values was acquired.
In the experimental conditions, we observed an SNR$_c$ varying from approximately $14$\,dB at a distance of $0.5$\,m to approximately $0$\,dB at the maximum operating distance of about $4$\,m. Furthtermore, the SNR$_j$ is approximately $40$\,dB, and the fraction of outliers varies from about 5$\%$ up to approximately 20$\%$.


The amplitude of the observed sawtooth waveform is approximately $10$\,ns as
can be seen in Fig. \ref{fig:sawtooth}, which is consistent with the nominal clock frequency of $100$\,MHz of the FPGA boards used.
Furthermore, the period is approximately 30 samples which, considering the sampling frequency of $5$\,kHz, corresponds to a frequency difference in the order of $200$\,Hz. Such a difference is consistent with the tolerance specification of the local oscillators used in the experiments.

The raw \textsc{RTT} data from the experimental measurement platform was directly used as the input for the WLS estimation algorithm. Whereas, the raw data was pre-processed by a coarse outlier-rejection method before being input into the ULS and PCP algorithms. In particular, the pre-processing outlier rejection method consists of the identification of the outliers by means of the same criterion in \eqref{e:outl_criterion}, and in the following substitution rule:
if the outlier is isolated, namely the preceding and following samples are not outliers, then a linear interpolation is performed. Specifically, the outlier sample is replaced by the average of the following and preceding sample. Conversely, if the outlier is not isolated, i.e. if the preceding or following samples are also outliers, the outlier sample is replaced by the median of the entire record.

For validation purposes, we also performed a characterization of ULS and PCP without using the pre-processing step. Results of such characterization, not presented here for brevity, show a considerable performance degradation of one order of magnitude or more with respect to WLS.

Therefore, in the following subsections, we provide results from experimental tests where the pre-processing method is applied to ULS and PCP.

\subsection{Frequency Difference Experimental Results}
We acquired \textsc{RTT} data from the system at four master-slave distances from $1$\,m to $4$\,m at $1$\,m steps. Simultaneously, the related ground-truth data, i.e. the true master-slave clock frequency difference, was acquired using the setup above and found to be approximately $-30$\,Hz with small variations among datasets at different distances.

\begin{figure}
 \centering
 \subfigure[]
   {\includegraphics[width = 0.9 \columnwidth]{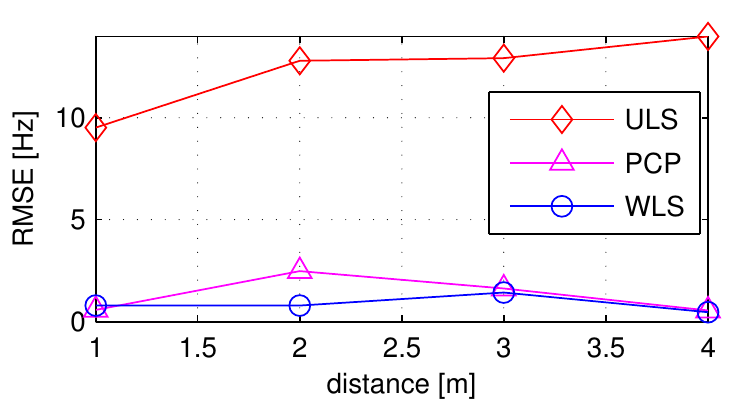}}
 \\
 \subfigure[]
   {\includegraphics[width = 0.9 \columnwidth]{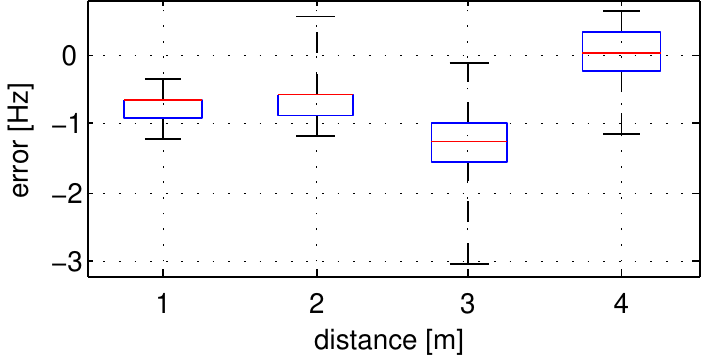}}
 \caption{Experimental results for the estimation of frequency difference. (a) RMSE of the frequency difference estimates vs range, for the three considered methods. (b) box plot of the frequency deviation estimation error vs range, for the proposed WLS estimator. The blue edges of each box denote the 25th and 75th percentiles, the central red mark denotes the median, and the top and bottom ends of the black dashed lines denote the maximum and minimum values, respectively.}
 \label{f:experim_results_fd}
\end{figure}

Subsequently, the acquired data was segmented into records, each consisting of 1000 consecutive samples. Each record was processed using the three considered methods. The resulting estimates were compared with the ground-truth, to obtain the results shown in Fig. \ref{f:experim_results_fd}. The proposed WLS method outperforms both the other considered methods, and its performance is relatively insensitive to variations in range. Moreover, the global RMSE, computed taking into account data acquired at all distances, is equal to $0.96$\,Hz for the proposed WLS estimator. Thus, the capability of sub-Hertz relative frequency measurement in experimental conditions is demonstrated.

\begin{figure}
\centering
\includegraphics[width = 0.95 \columnwidth]{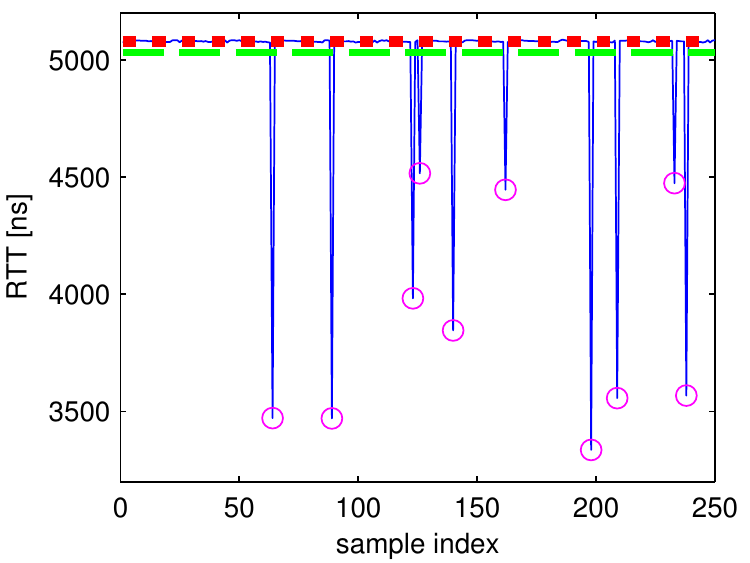}
\caption{Experimental data record showing outliers denoted with magenta circular markers. The data was obtained with the master and slave antennas placed 5 m apart. The dashed green line denotes the sample mean, which is clearly influenced by the outliers, and the dotted red line denotes the sample median which is robust against outliers.}
\label{f:data_outliers}
\end{figure}

Furthermore, a data record illustrating the presence of outliers is
shown in Fig.~\ref{f:data_outliers}. The presence of outliers can be
explained by analyzing the behavior of the receiver employed. In our experimental setup, in fact, the energy-detection receiver cannot distinguish between a proper response signal and an external RF interference signal. Therefore, when external interference bursts cross the pre-defined threshold in the energy detector, they cause spurious detections. Furthermore, the wideband noise which is inherently present in UWB systems can also cause spurious detections. Such spurious detections cause outliers during the \textsc{RTT} measurement procedure.

\subsection{Phase Experimental Results}
\begin{figure}
\centering
\includegraphics[width = 0.9 \columnwidth]{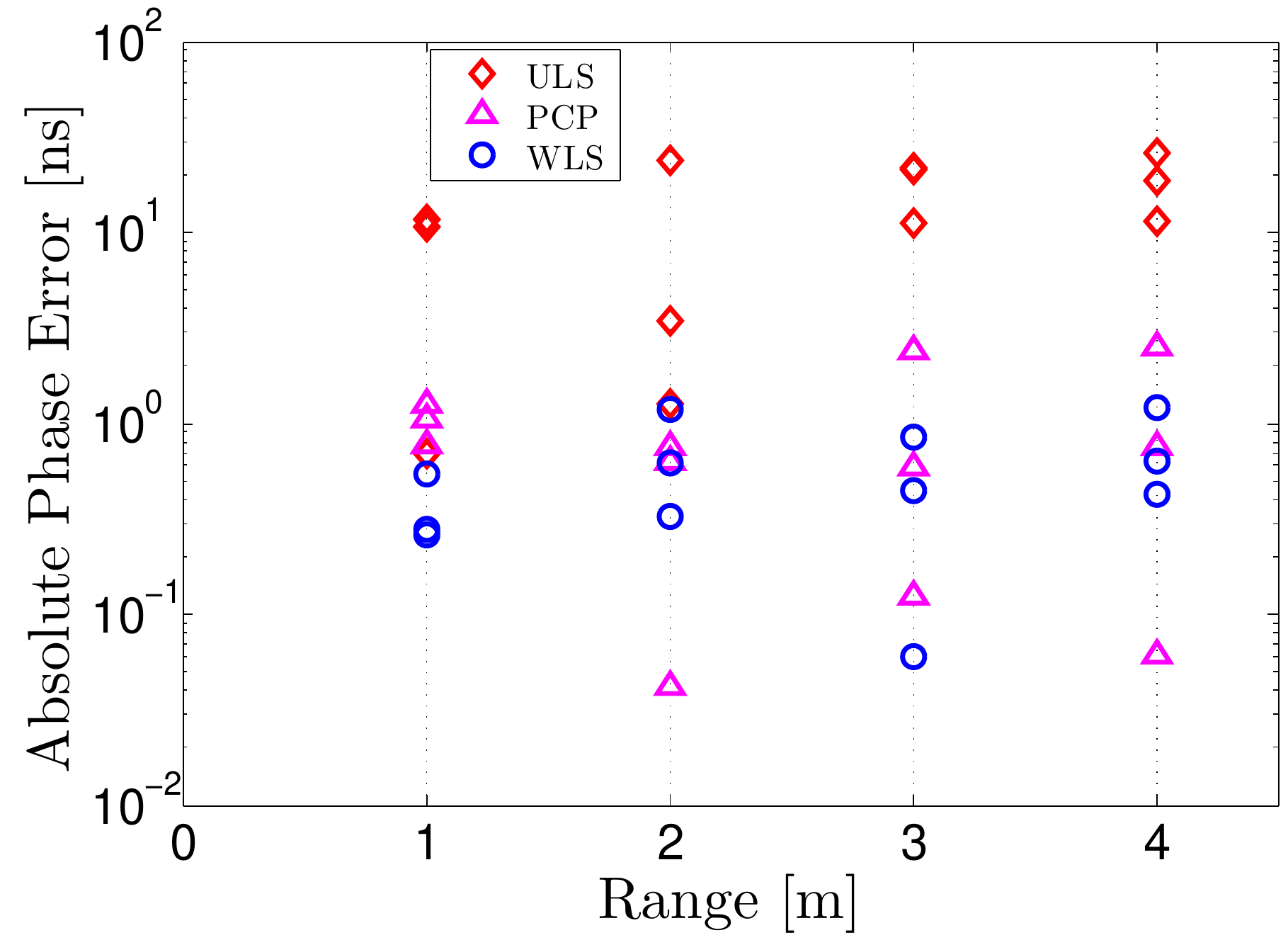}
\caption{Phase error measurements.}
\label{fig:phase_result}
\end{figure}

Figure \ref{fig:phase_result} shows error between measured phase and
estimated phase for various estimators.  Since the phase parameter is same for one set
of measurements, computing RMSE of phase estimate would require many set of
measurements which is practically difficult and time consuming. Hence,
instead of RMSE plots, a few phase error estimates are shown in the figure.
As can be seen, the phase estimation
error is of the order of $1$\,ns for WLS and PCP estimators. As expected
from simulation results, phase estimation error of ULS is very large.

\subsection{Range Experimental Results}
\begin{figure}
 \centering
 \subfigure[]
   {\includegraphics[width = 0.9 \columnwidth]{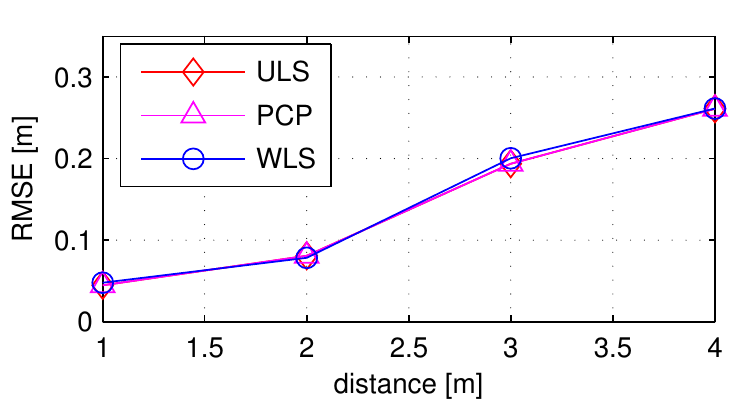}}
 \\
 \subfigure[]
   {\includegraphics[width = 0.9 \columnwidth]{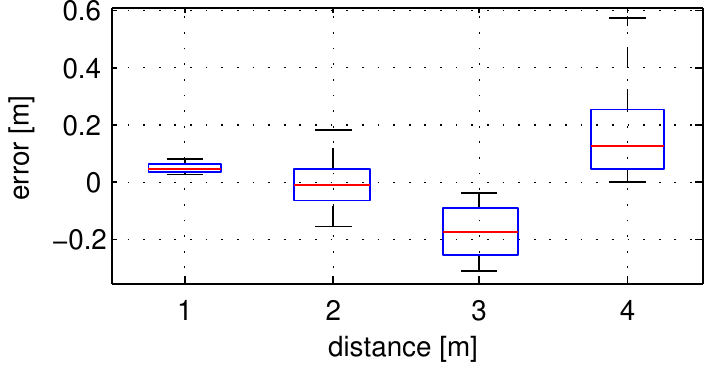}}
 \caption{Experimental results for the estimation of range. (a) RMSE of the range estimation error vs actual range, for the three considered methods. (b) box plot of the range estimation error vs actual range, for the proposed WLS estimator. The blue edges of each box denote the 25th and 75th percentiles, the central red mark denotes the median, and the top and bottom ends of the black dashed lines denote the maximum and minimum values, respectively. }
 \label{f:experim_results_range}
\end{figure}

In order to relate the measured \textsc{RTT} to the master-slave range, we performed
a calibration procedure based on that described in detail in
\cite{DeAngelisEtAl2013}. The calibration procedure removes various
systematic delays, particularly delay in RF front end of the
transceivers. In particular, 1000 \textsc{RTT} samples were acquired at a set of known distances from $0.5$\,m to $4.5$\,m at $0.5$\,m steps. Then, the calibration curve was calculated by applying a fifth-order polynomial fitting to the experimental data. The range estimate characterization was then performed on an independent dataset, acquired at four distances from $1$\,m to $4$\,m at $1$\,m steps. We denote the latter dataset as the \emph{validation} dataset. For each distance in the validation dataset, we acquired multiple 1000-sample records and processed each record using all the considered methods. The range estimation results are summarized in Fig. \ref{f:experim_results_range}. As clearly noticeable from Fig. \ref{f:experim_results_range}a, the difference between the three considered estimation methods is less than $1$\,cm and, therefore, negligible. The resulting global RMSE, which is calculated over the entire validation dataset, is $17$ \,cm for the proposed WLS method.

\subsection{Residual Analysis}\label{sec:residualanalysis}

\begin{figure}
\begin{center}

 \subfigure[]
   {\includegraphics[width = 0.9 \columnwidth]{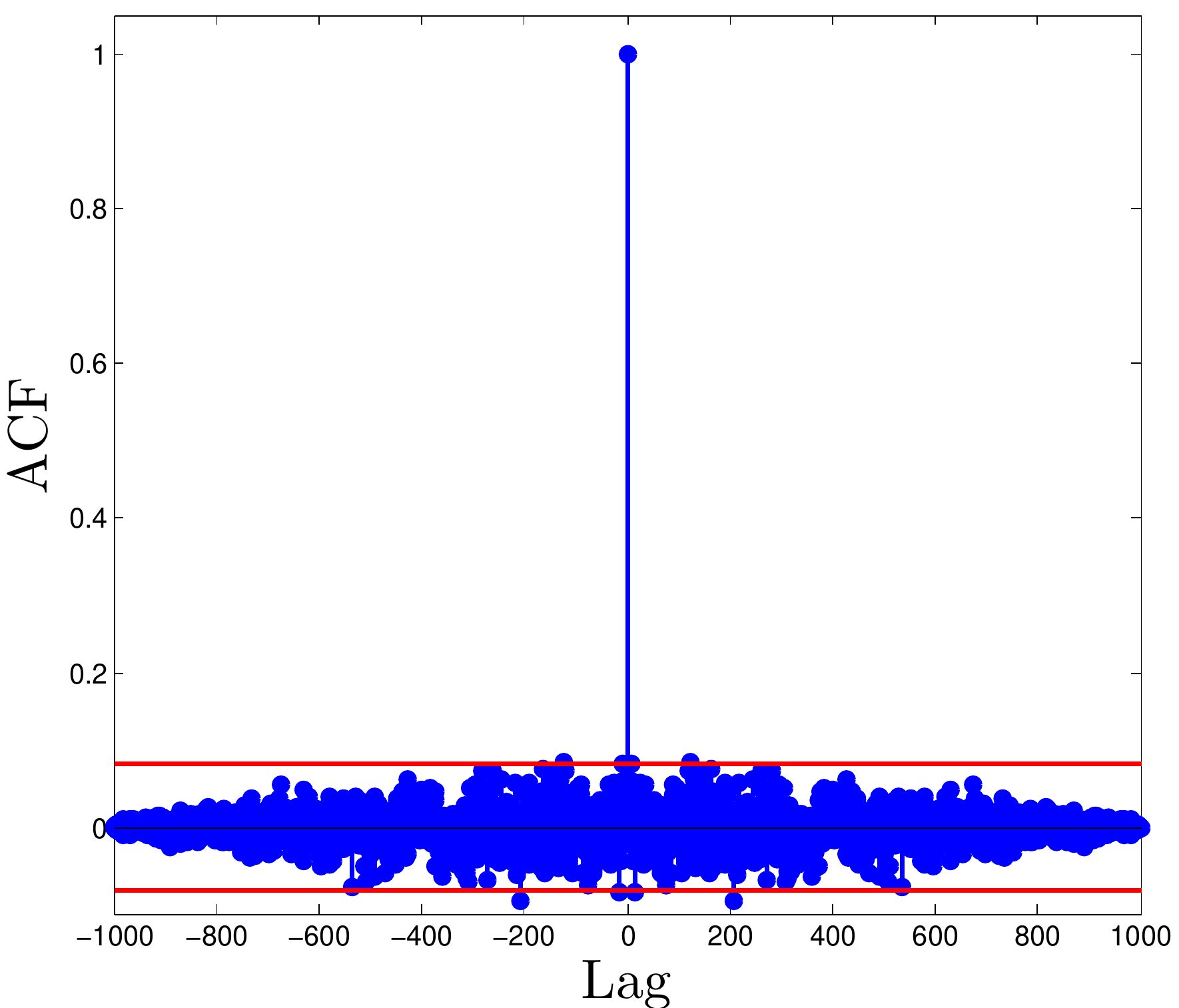}}
\end{center}
\caption{Autocorrelation function over the residuals of WLS, with 99\%
  confidence interval of a white noise sequence. Measurements at $4$\,m. }
\label{f:model_validation}
\end{figure}

Figure \ref{f:model_validation} shows the estimated autocorrelation
function over the residuals of the WLS estimator. The residuals are obtained
after subtracting noiseless signal template as in (\ref{eq:model_vectorreduced})  using
WLS estimates from the measured data at $4$\,m. The procedure of analyzing
residuals is similar to as discussed in
\cite{ref_resid}. Fig.~\ref{f:model_validation} also plots the
$99\%$ confidence bounds for a white noise sequence. As can
be seen, virtually all ACF points fall within these bounds except at
the zero lag which thus validates the model proposed in
(\ref{eq:model_vectorreduced}).

\section{Conclusion}\label{sec:conclusion}
In this paper we have proposed a new method to generate wireless measurements for
 joint clock parameter estimation and range estimation between the
 clocks of two nodes. We have formulated a mathematical model which captures the
 measurement generating phenomena. Further, we have developed two
 robust estimators (PCP and WLS) with different complexities and have
 compared their performance with the popular ULS estimator. The
 estimators were compared by simulations using the proposed
 model. We suggested outlier rejection techniques to handle real-world
 measurements and compared the estimator performances using data
 generated by a UWB testbed, reporting favorable experimental RMSE. 

 We envision that this method has a two-fold use: firstly as a building-block
in positioning systems which require wireless node-to-node synchronization. 
Secondly, this method will give rise to applications where accurate wireless clock
synchronization would be beneficial, such as in the internet-of-things,
and the machine-to-machine communication frameworks.


\begin{appendix}
To derive the WLS estimator in the efficient form of \eqref{e:WLS_estimator},
we first define the matrix 
\begin{equation*}
\begin{split}
\mbs{\Pi} \triangleq \mbf{1}
(\mbf{1}^\top \mbf{W} \mbf{1})^{-1} \mbf{1}^\top \mbf{W}  =  \frac{\mbf{1}\mbf{w}^\top}{\mbf{1}^\top \mbf{w}},
\end{split}
\end{equation*}
where the equality follows from $\mbf{W}\mbf{1} = \mbf{w}$. Next, recall that $\mbf{r}(f_d, \phi) =  \mbf{y} - h(f_d, \phi, \mbf{0})
- \delta_0 \mbf{1}$. Inserting \eqref{e:WLS_rho} back into
\eqref{eq:J_wls}, yields a concentrated cost function which we can
re-write:
\begin{equation*}
\begin{split}
J(f_d, \phi, \hat{\rho}) &= \| \mbf{r} - \mbs{\Pi}\mbf{r}
\|^2_{\mbf{W}} \\
&= \mbf{r}^\top \mbf{W}\mbf{r} - 2 \mbf{r}^\top \mbf{W}\mbs{\Pi}
\mbf{r} + \mbf{r}^\top \mbs{\Pi}^\top \mbf{W} \mbs{\Pi} \mbf{r} \\
&= \mbf{r}^\top \mbf{W}\mbf{r}  - \mbf{r}^\top \mbf{W}\mbs{\Pi}
\mbf{r} \\
&= \| \mbf{W}^{1/2} \mbf{r} \|^2_2 - \frac{ \mbf{r}^\top \mbf{W}
  \mbf{1}\mbf{w}^\top \mbf{r}}{\mbf{1}^\top \mbf{w}} \\
&=  \| \mbf{w}^{1/2} \odot \mbf{r} \|^2_2 - \frac{ |\mbf{w}^\top \mbf{r}|^2}{\mbf{1}^\top \mbf{w}} ,
\end{split}
\end{equation*}
using the fact that $\mbs{\Pi}^\top \mbf{W} \mbs{\Pi} =  \mbf{W}
\mbs{\Pi}$, thus reproducing \eqref{e:WLS_estimator}.

\end{appendix}

\bibliography{refs_clock_synch}

\begin{thebibliography}{10}

\bibitem{gubbi2013internet}
J.~Gubbi, R.~Buyya, S.~Marusic, and M.~Palaniswami, ``Internet of things
  (\textsc{IoT}): A vision, architectural elements, and future directions,''
  {\em Future Generation Computer Systems}, vol.~29, no.~7, pp.~1645--1660,
  2013.

\bibitem{M2M1}
S.-Y. Lien, K.-C. Chen, and Y.~Lin, ``Toward ubiquitous massive accesses in
  3gpp machine-to-machine communications,'' {\em Communications Magazine,
  IEEE}, vol.~49, pp.~66--74, April 2011.

\bibitem{M2M2}
Z.~Fadlullah, M.~Fouda, N.~Kato, A.~Takeuchi, N.~Iwasaki, and Y.~Nozaki,
  ``Toward intelligent machine-to-machine communications in smart grid,'' {\em
  Communications Magazine, IEEE}, vol.~49, pp.~60--65, April 2011.

\bibitem{ubisense_ref}
\textsc{Ubisense}, ``Ubisense \textsc{RTLS} system.''
  \url{http://www.ubisense.net/en/}, 2014.

\bibitem{zheng_wu}
J.~Zheng and Y.-C. Wu, ``Joint time synchronization and localization of an
  unknown node in wireless sensor networks,'' {\em Signal Processing, IEEE
  Transactions on}, vol.~58, pp.~1309--1320, March 2010.

\bibitem{DBP_joint}
B.~Denis, J.-B. Pierrot, and C.~Abou-Rjeily, ``Joint distributed
  synchronization and positioning in \textsc{UWB} ad hoc networks using toa,''
  {\em Microwave Theory and Techniques, IEEE Transactions on}, vol.~54,
  pp.~1896--1911, June 2006.

\bibitem{TDOA_clock_Gholami}
M.~Gholami, S.~Gezici, and E.~Strom, ``\textsc{TDOA} based positioning in the
  presence of unknown clock skew,'' {\em Communications, IEEE Transactions on},
  vol.~61, pp.~2522--2534, June 2013.

\bibitem{chepuri_2012}
S.~Chepuri, G.~Leus, and A.~van~der Veen, ``Joint localization and clock
  synchronization for wireless sensor networks,'' in {\em Signals, Systems and
  Computers (ASILOMAR), 2012 Conference Record of the Forty Sixth Asilomar
  Conference on}, pp.~1432--1436, Nov 2012.

\bibitem{chepuri_2013}
S.~Chepuri, R.~Rajan, G.~Leus, and A.-J. van~der Veen, ``Joint clock
  synchronization and ranging: Asymmetrical time-stamping and passive
  listening,'' {\em Signal Processing Letters, IEEE}, vol.~20, pp.~51--54, Jan
  2013.

\bibitem{etzlinger_2014}
B.~Etzlinger, F.~Meyer, H.~Wymeersch, F.~Hlawatsch, G.~Muller, and A.~Springer,
  ``Cooperative simultaneous localization and synchronization: Toward a
  low-cost hardware implementation,'' in {\em Sensor Array and Multichannel
  Signal Processing Workshop (SAM), 2014 IEEE 8th}, pp.~33--36, June 2014.

\bibitem{carroll_2014}
P.~Carroll, K.~Mahmood, S.~Zhou, H.~Zhou, X.~Xu, and J.-H. Cui, ``On-demand
  asynchronous localization for underwater sensor networks,'' {\em Signal
  Processing, IEEE Transactions on}, vol.~62, pp.~3337--3348, July 2014.

\bibitem{henk_positioning1}
H.~Wymeersch, J.~Lien, and M.~Win, ``Cooperative localization in wireless
  networks,'' {\em Proceedings of the IEEE}, vol.~97, pp.~427--450, Feb 2009.

\bibitem{molisch_pos_UWB}
S.~Gezici, Z.~Tian, G.~Giannakis, H.~Kobayashi, A.~Molisch, H.~Poor, and
  Z.~Sahinoglu, ``Localization via ultra-wideband radios: a look at positioning
  aspects for future sensor networks,'' {\em Signal Processing Magazine, IEEE},
  vol.~22, pp.~70--84, July 2005.

\bibitem{schedule_comm_lett}
S.~Dwivedi, D.~Zachariah, A.~De~Angelis, and P.~H{\"a}ndel, ``Cooperative
  decentralized localization using scheduled wireless transmissions,'' {\em
  Communications Letters, IEEE}, vol.~17, pp.~1240--1243, June 2013.

\bibitem{NTP_paper}
D.~Mills, ``Internet time synchronization: the network time protocol,'' {\em
  Communications, IEEE Transactions on}, vol.~39, pp.~1482--1493, Oct 1991.

\bibitem{ganeriwal2003timing}
S.~Ganeriwal, R.~Kumar, and M.~B. Srivastava, ``Timing-sync protocol for sensor
  networks,'' in {\em Proceedings of the 1st international conference on
  Embedded networked sensor systems}, pp.~138--149, ACM, 2003.

\bibitem{elson2002fine}
J.~Elson, L.~Girod, and D.~Estrin, ``Fine-grained network time synchronization
  using reference broadcasts,'' {\em ACM SIGOPS Operating Systems Review},
  vol.~36, no.~SI, pp.~147--163, 2002.

\bibitem{DeAngelisEtAl2013}
A.~De~Angelis, S.~Dwivedi, and P.~H\"andel, ``Characterization of a flexible
  {UWB} sensor for indoor localization,'' {\em IEEE Trans. Instrum. Meas.},
  vol.~62, pp.~905--913, may 2013.

\bibitem{Cui_scholtz}
C.-C. Chui and R.~Scholtz, ``Time transfer in impulse radio networks,'' {\em
  Communications, IEEE Transactions on}, vol.~57, pp.~2771--2781, September
  2009.

\bibitem{carbone_perugia}
P.~Carbone, A.~Cazzorla, P.~Ferrari, A.~Flammini, A.~Moschitta, S.~Rinaldi,
  T.~Sauter, and E.~Sisinni, ``Low complexity \textsc{UWB} radios for precise
  wireless sensor network synchronization,'' {\em Instrumentation and
  Measurement, IEEE Transactions on}, vol.~62, pp.~2538--2548, Sept 2013.

\bibitem{Serpedin_YUW}
Y.-C. Wu, Q.~Chaudhari, and E.~Serpedin, ``Clock synchronization of wireless
  sensor networks,'' {\em Signal Processing Magazine, IEEE}, vol.~28,
  pp.~124--138, Jan 2011.

\bibitem{Bernhard_Henk}
B.~Etzlinger, H.~Wymeersch, and A.~Springer, ``Cooperative synchronization in
  wireless networks,'' {\em Signal Processing, IEEE Transactions on}, vol.~62,
  pp.~2837--2849, June 2014.

\bibitem{Ferris_Kumar}
N.~Freris, S.~Graham, and P.~Kumar, ``Fundamental limits on synchronizing
  clocks over networks,'' {\em Automatic Control, IEEE Transactions on},
  vol.~56, pp.~1352--1364, June 2011.

\bibitem{Isaac_handel}
I.~Skog and P.~H{\"a}ndel, ``Synchronization by two-way message exchanges:
  Cram\'er-rao bounds, approximate maximum likelihood, and offshore submarine
  positioning,'' {\em Signal Processing, IEEE Transactions on}, vol.~58,
  pp.~2351--2362, April 2010.

\bibitem{schedule_self}
D.~Zachariah, A.~De~Angelis, S.~Dwivedi, and P.~H{\"a}ndel, ``Self-localization
  of asynchronous wireless nodes with parameter uncertainties,'' {\em Signal
  Processing Letters, IEEE}, vol.~20, pp.~551--554, June 2013.

\bibitem{schedule_eurasip}
D.~Zachariah, A.~Angelis, S.~Dwivedi, and P.~H{\"a}ndel, ``Schedule-based
  sequential localization in asynchronous wireless networks,'' {\em EURASIP
  Journal on Advances in Signal Processing}, vol.~2014, no.~1, 2014.

\bibitem{shang2003localization}
Y.~Shang, W.~Ruml, Y.~Zhang, and M.~P. Fromherz, ``Localization from mere
  connectivity,'' in {\em Proceedings of the 4th ACM international symposium on
  Mobile ad hoc networking \& computing}, pp.~201--212, ACM, 2003.

\bibitem{experiment_win_dardari}
A.~Conti, M.~Guerra, D.~Dardari, N.~Decarli, and M.~Win, ``Network
  experimentation for cooperative localization,'' {\em Selected Areas in
  Communications, IEEE Journal on}, vol.~30, pp.~467--475, February 2012.

\bibitem{tretter_paper}
S.~Tretter, ``Estimating the frequency of a noisy sinusoid by linear regression
  (corresp.),'' {\em Information Theory, IEEE Transactions on}, vol.~31,
  pp.~832--835, Nov 1985.

\bibitem{skog_handel2}
I.~Skog and P.~H{\"a}ndel, ``Time synchronization errors in loosely coupled
  gps-aided inertial navigation systems,'' {\em Intelligent Transportation
  Systems, IEEE Transactions on}, vol.~12, pp.~1014--1023, Dec 2011.

\bibitem{MSTINFOCOM}
S.~Moon, P.~Skelly, and D.~Towsley, ``Estimation and removal of clock skew from
  network delay measurements,'' in {\em INFOCOM '99. Eighteenth Annual Joint
  Conference of the IEEE Computer and Communications Societies. Proceedings.
  IEEE}, vol.~1, pp.~227--234 vol.1, Mar 1999.

\bibitem{acam_TDC}
\textsc{ACAM}, ``{\textsc{acam} - Solutions in Time}.''
  \url{http://www.acam.de/}, 2014.
\newblock [Online; accessed 17-June-2014].

\bibitem{volker2002frequency}
B.~Volker and P.~Handel, ``Frequency estimation from proper sets of
  correlations,'' {\em Signal Processing, IEEE Transactions on}, vol.~50,
  no.~4, pp.~791--802, 2002.

\bibitem{matlab_unwrap}
\textsc{Mathworks}, ``{Matlab 'unwrap' function}.''
  \url{http://www.mathworks.se/help/matlab/ref/unwrap.html}, 2014.
\newblock [Online; accessed 17-June-2014].

\bibitem{ZoubirEtAl2012}
A.~Zoubir, V.~Koivunen, Y.~Chakhchoukh, and M.~Muma, ``Robust estimation in
  signal processing: A tutorial-style treatment of fundamental concepts,'' {\em
  Signal Processing Magazine, IEEE}, vol.~29, no.~4, pp.~61--80, 2012.

\bibitem{Stoica&Moses2005_spectral}
P.~Stoica and R.~L. Moses, {\em Spectral analysis of signals}.
\newblock Pearson/Prentice Hall Upper Saddle River, NJ, 2005.

\bibitem{rousseeuw1984least}
P.~J. Rousseeuw, ``Least median of squares regression,'' {\em Journal of the
  American statistical association}, vol.~79, no.~388, pp.~871--880, 1984.

\bibitem{Hodge&Austin2004}
V.~J. Hodge and J.~Austin, ``A survey of outlier detection methodologies,''
  {\em Artificial Intelligence Review}, vol.~22, no.~2, pp.~85--126, 2004.

\bibitem{jitter_ref}
T.~Neu, ``Clock jitter analyzed in the time domain, part 1.''
  http://www.ti.com/lit/an/slyt379/slyt379.pdf, 2010.

\bibitem{Xilinx_ML505}
Xilinx, ``Ug347 ml505 evaluation platform, user guide.''
  \url{http://www.xilinx.com/support/documentation/boards\_and\_kits/ug347.pdf},
  2011.

\bibitem{laser_ref}
\textsc{FLUKE}, ``Laser distance meter.''
  \url{http://www.fluke.com/fluke/auen/laser-distance-meters/fluke-411d-laser-distance-meter.htm?PID=69331},
  2014.

\bibitem{agilent_counter}
\textsc{Agilent}, ``{53230A 350 MHz Universal Frequency Counter/Timer, 12
  digits/s, 20 ps}.''
  \url{http://www.home.agilent.com/en/pd-1893420-pn-53230A/350-mhz-universal-frequency-counter-timer-12-digits-s-20-ps?&cc=SE&lc=eng},
  2014.
\newblock [Online; accessed 17-June-2014].

\bibitem{Satyam_ISPCS}
S.~Dwivedi and P.~H{\"a}ndel, ``Precise clock parameter estimation and ground
  truth capture for clock error measurements using \textsc{FPGA}s,'' in {\em
  International IEEE Symposium on Precision Clock Synchronization for
  Measurement, Control and Communication (ISPCS)}, (Austin, Tx), September
  2014.

\bibitem{ref_resid}
Math\textsc{W}orks, ``{Residual Analysis with Autocorrelation}.''
  \url{http://www.mathworks.se/help/signal/ug/residual-analysis-with-autocorrelation.html},
  2014.
\newblock [Online; accessed 17-June-2014].

\end{thebibliography}
\bibliographystyle{ieeetr}


\end{document}